\documentclass[prd, aps, nofootinbib]{revtex4}

\newcommand{\checked}[1]{}

\usepackage{float}
\usepackage{epsfig}
\usepackage{color}
\usepackage{tabularx}
\usepackage{slashbox}
\usepackage{diagbox}
\usepackage{amsmath}
\usepackage{graphicx}
\usepackage{booktabs}
\usepackage{hyperref}
\hypersetup{
    colorlinks=true,
    linkcolor=blue,
    citecolor=blue,
    filecolor=blue,
    urlcolor=blue,
}

\newcommand{\be}{\begin{equation}}
\newcommand{\ee}{\end{equation}}
\newcommand{\ba}{\begin{eqnarray}}
\newcommand{\ea}{\end{eqnarray}}

\newcommand{\la}{\label}
\newcommand{\sig}{\sigma}

\begin{document}

\title {Effective Debye Screening Mass in an Anisotropic Quark Gluon Plasma}

\author{Lihua Dong$^{a,b,c}$, Yun Guo$^{a,b,}$\footnote{yunguo@mailbox.gxnu.edu.cn}, Ajaharul Islam$^{d}$, and Michael Strickland$^{d,}$\footnote{mstrick6@kent.edu}}
\affiliation{
$^a$ Department of Physics, Guangxi Normal University, Guilin, 541004, China\\
$^b$ Guangxi Key Laboratory of Nuclear Physics and Technology, Guilin, 541004, China\\
$^c$ School of Physics and Astronomy, Sun Yat-Sen University, Zhuhai, 519082, China\\
$^d$ Department of Physics, Kent State University, Kent, OH 44242, United States\\}

\begin{abstract}
Due to the rapid longitudinal expansion of the quark-gluon plasma created in heavy-ion collisions, large local-rest-frame momentum-space anisotropies are generated during the system's evolution.  These momentum-space anisotropies complicate the modeling of heavy-quarkonium dynamics in the quark-gluon plasma due to the fact that the resulting inter-quark potentials are spatially anisotropic, requiring real-time solution of the 3D Schr\"odinger equation.  Herein, we introduce a method for reducing anisotropic heavy-quark potentials to isotropic ones by introducing an effective screening mass that depends on the quantum numbers $l$ and $m$ of a given state.  We demonstrate that, using the resulting effective Debye screening masses, one can solve a 1D Schr\"odinger equation and reproduce the full 3D results for the energies and binding energies of low-lying heavy-quarkonium bound states to relatively high accuracy.  The resulting effective isotropic potential models could provide an efficient method for including momentum-anisotropy effects in open quantum system simulations of heavy-quarkonium dynamics in the quark-gluon plasma.
\end{abstract}

\maketitle

\section{Introduction}\la{intro}

The ongoing heavy-ion collision experiments at the Relativistic Heavy Ion Collider and the Large Hadron Collider aim to create and study a primordial state of matter called a quark-gluon plasma (QGP).  One of the key observables that experimentalists are interested in is the production of heavy-quarkonium bound states in nucleus-nucleus (AA) collisions relative to their production in proton-proton (pp) collisions.  This observable is of interest because it is expected that the formation of heavy-quarkonium bound states is suppressed in AA collisions relative to that in pp collisions.  In early models the suppression of heavy-quarkonium bound states was solely due to the Debye screening of the inter-quark potential  \cite{Matsui:1986dk,Karsch:1987pv,Shuryak:2004tx,Shuryak:2003ty}.  Since these early works, it has been shown that it is equally important to include effects of in-medium singlet-octet transitions and Landau damping of the exchanged gluons, which results in an imaginary contribution to the heavy-quark (HQ) potential \cite{Laine:2006ns,Dumitru:2007hy,Brambilla:2008cx,Burnier:2009yu,Dumitru:2009fy,Dumitru:2009ni,Margotta:2011ta,Du:2016wdx,Nopoush:2017zbu,Guo:2018vwy}.  These effects can be modelled systematically using numerical solutions of the Lindblad equation which governs the evolution of the in-medium heavy-quarkonium reduced density matrix \cite{Brambilla:2017zei,Brambilla:2020qwo,Brambilla:2021wkt,Omar:2021kra}.

One of the limitations of prior studies of the evolution of the heavy-quarkonium reduced density matrix is that they explicitly rely on an assumption of momentum-space isotropy.  This assumption allows one to simplify the resulting dynamical evolution equations such that only solution of a one-dimensional Schr\"odinger equation (SE) subject to stochastic jumps is required \cite{Brambilla:2021wkt}.  This assumption can, however, not be made in a real-world quark-gluon plasma due to the rapid longitudinal expansion of the QGP and the existence of a finite relaxation time of the system.  As a result, the QGP develops a high degree of momentum anisotropy in its local rest frame, see e.g. \cite{Strickland:2014pga,Berges:2020fwq} and references therein.  This complication means that, in practice, one must solve a 3D stochastic Schr\"odinger equation or corresponding Lindblad equation, which is numerically prohibitive.  In this paper, we introduce a method to obtain effective 1D isotropic potentials from underlying anisotropic potentials.  The method introduced can be used to simplify the inclusion of momentum-space anisotropy effects on heavy-quarkonium dynamics.

In order to develop and test this method, we will make use of a widely used anisotropic distribution function ansatz called the Romatschke-Strickland (RS) form~\cite{Romatschke:2003ms,Romatschke:2004jh}
\be\label{anisodis}
f_{\rm aniso}({\bf k})\equiv f_{\rm iso}\!\left(\frac{1}{\lambda}\sqrt{{\bf k}^2+\xi ({\bf k}\cdot {\bf n})^2}\right) ,
\ee
\checked{m}
where the isotropic distribution function $f_{\rm iso}$ is an arbitrary isotropic distribution function that decreases sufficiently rapidly at large momentum. In the above equation, the unit vector $\bf{n}$ is used to denote the direction of anisotropy and $\lambda$ is a temperature-like scale which, in the thermal equilibrium limit, should be understood as the temperature $T$ of the system. In addition, we use an adjustable parameter $\xi$ to quantify the degree of momentum-space anisotropy
\begin{equation}
\xi = \frac{1}{2}\frac{\langle \bf k^{2}_{\perp}\rangle}{\langle k^2_z\rangle}-1~,
\end{equation}
where $k_z \equiv \bf k \cdot n$ and $\bf k_{\perp}\equiv \bf k-\bf n(k\cdot n)$ correspond to the particle momenta along and perpendicular to the direction of anisotropy, respectively. By assuming $\bf{n}$ to be parallel to the beam-line direction, the case $\xi > 0$ 
is relevant to high-energy heavy-ion experiments.

Among the many new phenomena which emerge as a consequence of QGP momentum-space anisotropy, herein we are interested in the in-medium properties of quarkonium states. In the non-relativistic limit, the binding energies and the decay widths of the bound states can be obtained by solving the 3D Schr\"odinger equation with a specified heavy-quark potential which describes the force between the quark and antiquark. Given the distribution function in Eq.~(\ref{anisodis}), the complex-valued potential has been computed in the resummed hard-thermal-loop (HTL) perturbation theory~\cite{Dumitru:2007hy,Guo:2008da,Dumitru:2009fy}.\footnote{We consider $f_{\rm iso}$ in thermal equilibrium which is simply the Fermi-Dirac or Bose-Einstein distribution function.} For arbitrary anisotropy $\xi$, the real part of the potential has to be evaluated numerically and analytical results can only be obtained in the limits, e.g., $|\xi| \ll 1$, $\xi \gg 1$, and $\xi \rightarrow -1$. On the other hand, the determination of the imaginary part for general $\xi$ has encountered difficulties related to the presence of a pinch singularity in the static gluon propagator \cite{Nopoush:2017zbu}. As a result, this is still an open question and we will focus on the real part of the potential in the current work.

It is worth noting that the perturbative HQ potential in an anisotropic plasma can be well fitted with a standard Debye-screened function $\sim e^{-r \tilde{m}_D(\lambda,\xi,\theta)}/r$, therefore, is formally analogous to its counterpart in equilibrium \cite{Strickland:2011aa}. The determination of $\tilde{m}_D(\lambda,\xi,\theta)$ requires a specific angle $\theta$ denoting the quark pair alignment ${\bf r}$ with respect to the direction of anisotropy ${\bf n}$. However, such a $\theta$-dependence results in an unclear physical interpretation of the screening mass and only makes sense in a classical picture where the quark pair has a fixed alignment. In addition, solving the three-dimensional SE makes the numerical determination of the binding energies of quarkonium states much more complicated compared to the case where a spherically symmetric HQ potential can be used. At present, this is the main obstacle to developing phenomenological applications which include momentum-anisotropy effects.

In this work, we propose an angle-averaged screening mass ${\cal{M}}_{l m}(\lambda,\xi)$ in an anisotropic medium through the following definition
\ba\la{effm0}
{\cal{M}}_{l m}(\lambda,\xi)&=&\langle {\rm{Y}}_{l m}(\theta,\phi)| \tilde{m}_D(\lambda,\xi,\theta) | {\rm{Y}}_{l m}(\theta,\phi)\rangle\, ,\nonumber \\
&=&\int_{-1}^{1} d \cos \theta \int_{0}^{2\pi} d \phi {\rm{Y}}_{l m}(\theta,\phi)  \tilde{m}_D(\lambda,\xi,\theta) {\rm{Y}}^*_{l m}(\theta,\phi)\, ,
\ea
where ${\rm{Y}}_{l m}$ are spherical harmonics with the azimuthal quantum number $l$ and magnetic quantum number $m$. Notice that although no angular dependence appears in ${\cal{M}}_{l m}(\lambda,\xi)$, this quantity is not universal for all the states labelled by the set of quantum numbers $l$ and $m$. As will be discussed in this work, information on the physical properties of quarkonium states in an anisotropic QCD plasma can be obtained at a quantitative level by analyzing the corresponding problem in an ``isotropic" medium characterized by the angle-averaged screening mass ${\cal{M}}_{l m}(\lambda,\xi)$. In this sense, ${\cal{M}}_{l m}(\lambda,\xi)$ is also called the {\em effective screening mass}.

The rest of the paper is organized as the following. In Sec.~\ref{model}, we introduce the model construction of the HQ potential in an anisotropic QCD plasma and qualitatively explain the rationality of our definition of the {\em effective screening mass} as given by Eq.~(\ref{effm0}). In Sec.~\ref{pert}, an anisotropic HQ potential as given in Eq.~(\ref{kms}) is Taylor expanded around an isotropic configuration $\tilde{m}_D (\lambda,\xi,\theta)= {\cal{M}}_{l m}(\lambda, \xi)$ where the values of $l$ and $m$ are specified based on the concept of the ``most similar state". In Sec.~\ref{suppression}, we demonstrate that highly accurate results of the eigen/binding energies of quarkonium states can be obtained by solving the Schr\"odinger equation with only the leading order contribution in the above Taylor expansion of the HQ potential. In addition, an estimation on the discrepancy resulting from replacing $\tilde{m}_D (\lambda,\xi,\theta)$ with ${\cal{M}}_{l m}(\lambda, \xi)$ in Eq.~(\ref{kms}) is also obtained. Some applications of the {\em effective screening mass} derived herein are considered in Sec.~\ref{apps}. We briefly summarize our findings and give an outlook for the future in Sec.~\ref{conclusions}. Finally, for several low-lying heavy-quarkonium bound states, the exact 3D results of the eigen/binding energies based on the potential model in Eq.~(\ref{kms}), together with the corresponding discrepancies from using the one-dimensional potential model based on the {\em effective screening masses} are listed in Appendix~\ref{appa}, which provides a direct numerical check of our method.

\section{The heavy-quark potential in an anisotropic medium with effective screening mass}\la{model}

Previous works have demonstrated that the non-perturbative contributions to the HQ potential are not negligible for practical applications, therefore, the potential derived from perturbation theory can not fully capture the in-medium properties of quarkonium states. In general, describing the long-distance interaction between the quark pair relies on the phenomenological potential models which are constructed based on the lattice simulations. As a popular research topic, continuous attention has been paid on modeling the real part of the potential. In addition, developing a complex potential model has also been considered in recent years. However, most of the studies on quarkonium physics were limited to a momentum-space isotropic quark-gluon plasma. Due to the absence of the lattice simulations for a non-ideal anisotropic system, the corresponding potential models have to be introduced in an indirect way.

A first attempt to model the heavy-quark potential in an anisotropic plasma was carried out in Ref.~\cite{Dumitru:2009ni}. As a ``minimal" extension of the isotropic Karsch-Mehr-Satz (KMS) potential \cite{Matsui:1986dk,Karsch:1987pv} based on the internal energy, the anisotropic version was obtained by replacing the ideal Debye mass $m_D(\lambda)$ in thermal equilibrium with an anisotropic screening scale $\tilde{m}_D(\lambda,\xi,\theta)$. Explicitly, one assumes the following form for the (real part of the) potential model
\be\la{kms}
V(r,\tilde{m}_D)=-\frac{\alpha}{r}(1+\tilde{m}_D r)e^{-\tilde{m}_D r}+\frac{2\sig}{\tilde{m}_D}(1-e^{-\tilde{m}_D r})-\sig r e^{-\tilde{m}_D r}\, ,
\ee
where $\alpha$ is an effective Coulomb coupling at (moderately) short distance and $\sig$ is the string tension. In the rest of the paper, we choose $\alpha=C_F \alpha_s=0.385$ and $\sig=0.223\, {\rm GeV}^2$ for numerical evaluations. These two parameters are assumed to be unchanged in a hot medium, once determined at zero temperature. Following Ref.~\cite{Dumitru:2009ni}, the isotropic Debye mass is given by $m_D=A g \lambda \sqrt{1+N_f/6}$ with $N_f=2$ and $g=1.72$. The factor $A=1.4$ which accounts for non-perturbative effects has been determined in lattice calculations. In addition, we assume the critical temperature $T_c=192\ {\rm MeV}$.

For our purposes the most important feature of the KMS potential model is that medium effects are entirely encoded in the Debye mass $m_D(\lambda)$ and the very same screening scale that appears in the perturbative Coulombic term also shows up in the non-perturbative string contributions. Extending this assumption to anisotropic plasmas, the feasibility of the ``minimal" extension depends on the extraction of an angle-dependent screening mass $\tilde{m}_D(\lambda,\xi,\theta)$ from the perturbative potential which has been computed from the first principles. Despite this choice of potential, we emphasize that the generalization from an isotropic HQ potential to the anisotropic one is not restricted to the use of the KMS potential model. In fact, any other isotropic model, as long it shares this feature with the KMS model, can be generalized in a similar way. Some recent examples of such HQ potential models can be found in Refs.~\cite{Thakur:2013nia,Burnier:2015nsa,Krouppa:2017jlg,Guo:2018vwy,Lafferty:2019jpr}. For definiteness, in the following discussions, we will consider the potential model as given in Eq.~(\ref{kms}).

From a phenomenological point of view, we expect that non-zero anisotropy corrections only amount to a modification on the isotropic Debye mass $m_D(\lambda)$. In fact, this turns to be reasonable based on an analysis of the perturbative contributions in the HQ potential. In principle, the anisotropic screening mass $\tilde{m}_D(\lambda,\xi,\theta)$ can be defined as the inverse of the distance scale over which $|r V(r)|$ drops by a factor of $e$. For small anisotropy $\xi$, the screening mass is given by~\cite{Dumitru:2009ni}
\be\label{anisomd}
\tilde{m}_D(\lambda, \xi\ll 1, \theta)=m_D(\lambda) \bigg(1-\xi \frac{3+\cos 2\theta}{16}\bigg)\, .
\ee
As expected, the non-ideal Debye mass $\tilde{m}_D(\lambda,\xi,\theta)$ depends not only on the anisotropy parameter $\xi$ but also on the angle $\theta$ between ${\bf r}$ and ${\bf n}$. With this in hand, we can approximate the anisotropic potential at short distances with a Debye-screened Coulomb potential where the $\theta$-dependent screening mass is given by Eq.~(\ref{anisomd}). More interestingly, it is found that for arbitrary anisotropies, the perturbative potential can also be perfectly parameterized by using the same Debye-screened Coulomb potential although in this case, the anisotropic Debye mass $\tilde{m}_D(\lambda,\xi,\theta)$ takes a much more complicated form as~\cite{Strickland:2011aa}
\be\la{eff11}
\tilde{m}_D(\lambda,\xi,\theta)=m_D(\lambda) [f_1(\xi) \cos^2 \theta+f_2(\xi)]^{-\frac{1}{4}}\, ,
\ee
where
\ba
f_1(\xi)&=& \frac{9\xi (1+\xi)^\frac{3}{2}}{2\sqrt{3+\xi}(3+\xi^2)}\frac{\pi^2\big(\sqrt{2}-(1+\xi)^\frac{1}{8}\big)+16\big(\sqrt{3+\xi}-\sqrt{2}\big)}{(\sqrt{6}-\sqrt{3})\pi^2-16(\sqrt{6}-3)}\, , \nonumber \\
f_2(\xi) &=& \xi \Big(\frac{16}{\pi^2}-\frac{\sqrt{2}(16/\pi^2-1)+(1+\xi)^\frac{1}{8}}{\sqrt{3+\xi}}\Big) \Big(1-\frac{(1+\xi)^\frac{3}{2}}{1+\xi^2/3}\Big)+f_1(\xi)+1 \, .
\ea
The above expression ensures the correct asymptotic limits for small and large $\xi$ and efficiently reproduces the short-range anisotropic potential when compared to the direct numerical evaluation of the full result \cite{Strickland:2011aa}. The existence of such a parameterization further guarantees the justification of the model construction in an anisotropic medium as proposed above. Namely, the behaviors of the potential at short distances can be well described by the Debye-screened Coulomb potential from which an anisotropic screening mass can be extracted.

Although it is not surprising that a $\theta$-dependence emerges in the anisotropic screening mass, one has to face the technical difficulty of solving a two- or three-dimensional SE. On the other hand, if we replace $\tilde{m}_D(\lambda,\xi,\theta)$ with the {\em effective screening mass} as introduced in Eq.~(\ref{effm0}), the restoration of the spherical symmetry in the HQ potential greatly simplifies the numerical evaluations as one only needs to solve a one-dimensional problem. We believe this is very important for many studies concerning quarkonium physics in an anisotropic situation. Therefore, the key issue is to demonstrate the rationality of the definition of ${\cal{M}}_{l m}(\lambda,\xi)$ introduced in Eq.~\eqref{effm0}. Qualitatively, it can be understood by analyzing both the anisotropy and quantum number dependence of the {\em effective screening mass}.

Using Eq.~(\ref{anisomd}), for the anisotropic screening mass in small $\xi$ limit, the integrations in Eq.~(\ref{effm0}) can be performed analytically, leading to
\be\la{effm0small}
{\cal{M}}_{l m}(\lambda,\xi<1)=m_D(\lambda)\Big[1-\frac{\xi}{8}k(l,m)\Big]\, ,
\ee
with
\be
k(l,m)=\frac{6l(l+1)-2(m^2+2)}{4l(l+1)-3} \,.\label{klm}
\ee
As expected, the {\em effective screening mass} decreases with increasing anisotropy $\xi$ and a reduced screening effect exists in an anisotropic plasma since the values of ${\cal{M}}_{l m}(\lambda,\xi)$ are always smaller than the ideal Debye mass $m_D(\lambda)$. Furthermore, the observed polarization of quarkonium states with non-zero angular momentum in an anisotropic plasma can be qualitatively explained by the corresponding $m$-dependence as given in Eq.~(\ref{effm0small}). We find that instead of $m_D(\lambda)(1-\xi/5)$ for $P_0$ states, ${\cal{M}}_{l m}(\lambda,\xi)$ takes a different value $m_D(\lambda)(1-3 \xi /20)$ for $P_{\pm 1}$ states. As a result, one can expect that the first excited states of bottomonia\footnote{They are the $1P$ states of bottomonium, identified with the $\chi_b$. In an isotropic medium, there is no preferred polarization of the $\chi_b$ between states with different magnetic quantum number $m$. Degeneracy is removed in an anisotropic medium where $\chi_{b0}$ and $\chi_{b\pm1}$ correspond to $L_z=0$ and $L_z=\pm1$, respectively. } with $L_z=0$ have a smaller screening mass therefore is bounded more tightly as compared to those with $L_z={\pm 1}$. For arbitrary $\xi$, the above discussions also hold although in this case, the values of ${\cal{M}}_{l m}(\lambda,\xi)$ need to be determined numerically. In addition, since the function $k(l,m)$ is in the range of $1\le k(l,m) \le 8/5$, Eq.~(\ref{effm0small}) suggests that at a given temperature-like scale $\lambda$ and anisotropy $\xi$, the {\em effective screening mass} has a minimum value $m_D(\lambda)(1-\xi/5)$ with $l=1$ and $m=0$ while its maximum, equaling to $m_D(\lambda)(1-\xi/8)$,  is reached when $l\rightarrow \infty$ and $m=\pm l$. Numerical evaluations on the values of  ${\cal{M}}_{l m}(\lambda,\xi)$ for arbitrary $\xi$ turn to support the same conclusion and we give the corresponding results with anisotropies $\xi=5$ and $\xi=10$ for different $l$ and $m$ in Table~\ref{emD5} and Table~\ref{emD10}, respectively.
\\
\\

\begin{minipage}{\textwidth}
\begin{minipage}[t]{0.47\textwidth}
\centering
\makeatletter\def\@captype{table}\makeatother
\begin{tabular}{ |c|  c|  c|  c| c|  c|}
\hline
     \backslashbox{$m$}{$l$} & $     0    $  &  $    1   $  &  $     2    $  &  $    3     $  &  $    10    $    \\ \hline
	  0  & $ 0.6215 $    &    $0.5860$  &    $ 0.6007 $  &    $ 0.6001 $  &  $ 0.6012 $    \\ \hline
	  1  & $\backslash$  &    $0.6393$  &    $ 0.6058 $  &    $ 0.6048 $  &  $ 0.6017 $    \\ \hline
	  2  & $\backslash$  &  $\backslash$&    $ 0.6477 $  &    $ 0.6180 $  &  $ 0.6030 $    \\ \hline	
	  3  & $\backslash$  &  $\backslash$&  $\backslash$  &    $ 0.6526 $  &  $ 0.6053 $    \\ \hline
	 10  & $\backslash$  &  $\backslash$&  $\backslash$  &  $\backslash$  &  $ 0.6642 $    \\ \hline	
\end{tabular}
\caption{The ratio ${\cal{M}}_{l m}/m_D$ at $\xi=5$.}
\label{emD5}
\end{minipage}
\begin{minipage}[t]{0.47\textwidth}
\centering
\makeatletter\def\@captype{table}\makeatother
\begin{tabular}{|c|  c|  c|  c| c|  c|}
\hline
      \backslashbox{$m$}{$l$}& $     0    $  &  $    1   $  &  $     2    $  &  $    3     $  &  $    10    $    \\ \hline
	    0  & $ 0.5242 $  &    $0.5001$  &    $ 0.5095 $  &    $ 0.5094 $  &  $ 0.5102 $    \\ \hline
	  1  & $\backslash$  &    $0.5363$  &    $ 0.5140 $  &    $ 0.5127 $  &  $ 0.5105 $    \\ \hline
	  2  & $\backslash$  &  $\backslash$&    $ 0.5419 $  &    $ 0.5223 $  &  $ 0.5115 $    \\ \hline	
	  3  & $\backslash$  &  $\backslash$&  $\backslash$  &    $ 0.5451 $  &  $ 0.5131 $    \\ \hline
	 10  & $\backslash$  &  $\backslash$&  $\backslash$  &  $\backslash$  &  $ 0.5527 $    \\ \hline		
\end{tabular}
\caption{The ratio ${\cal{M}}_{l m}/m_D$ at $\xi=10$.}
\label{emD10}
\end{minipage}
\end{minipage}

\section{Perturbative evaluations on the energies of quarkonium states}
\la{pert}

The above discussions suggest that it is reasonable to introduce an angle-averaged screening mass as defined in Eq.~(\ref{effm0}) to replace the angle-dependent mass in Eq.~(\ref{kms}). However, the discrepancy resulting from such a replacement has to be estimated at a quantitative level to ensure an accurate determination on the physical properties of quarkonium states in an anisotropic plasma. We start with the stationary Sch\"ordinger equation
\be\la{sse}
-\frac{1}{2 m_Q}\Big(\frac{1}{r^2}\frac{\partial}{\partial r}r^2\frac{\partial}{\partial r}-\frac{1}{r^2}{\hat L}^2\Big)\psi({\bf r})=[E-V({\bf r})]\psi({\bf r})\, ,
\ee
where $m_Q$ is the reduced mass for the quarkonium bound state with eigen-energy $E$ and ${\hat L}^2$ is the square of the angular-momentum operator. For an isotropic potential $V(r)$ without the dependence on the azimuthal angle $\theta$ and polar angle $\phi$, the wave-function $\psi({\bf r})$ can be separated into a radial and an angular part as
\be\la{isowf}
\psi^{\rm iso}_{nlm}({\bf r})=\frac{R_{nl}(r)}{r}Y_{lm}(\theta,\phi)\, ,
\ee
and the associated eigen-energy denoted by $E_{nl}$ is determined by the following equation
\be\label{raeq}
\Big[\frac{1}{2m}\Big(-\frac{{\rm d^2}}{{\rm d} r^2}+\frac{l(l+1)}{r^2} \Big)+ V(r)\Big]R_{nl}(r)=E_{nl} R_{nl}(r)\, .
\ee
In general, Eq.~(\ref{raeq}) can not be solved analytically but the numerical evaluations can be simply carried out. According to the well-known nodal theorem, the level of excitations can be identified by the number $n$ of the nodes of the reduced radial wave-function $R_{nl}(r)$. For a given $l$, the ground-state wave-function corresponds to $n=0$, while the wave-function of some radial excitations has a non-zero and finite number of nodes. Therefore, the number $n=0,1,2,3\cdots$. In addition, the spherical harmonics $Y_{lm}(\theta,\phi)$ in Eq.~(\ref{isowf}) is standard with $m=-l,-l+1,\cdots, 0,\cdots, l-1, l$ and $l=0,1,2,3,\cdots$.

Because of the breaking of the spherical symmetry, for an anisotropic potential $V({\bf r})$, the above separation of the wave-function is no longer true, namely, $\psi^{\rm iso}_{nlm}({\bf r})$ is not an eigen-state for the system.  On the other hand, due to the completeness of the eigen-functions in Eq.~(\ref{isowf}), we can expand $\psi_{\rm aniso}({\bf r})$ as the following
\be\la{anwf}
\psi_{\rm aniso}^{[k]}({\bf r})=\sum_{nlm} C^{[k]}_{nlm}\frac{R_{nl}(r)}{r}Y_{lm}(\theta,\phi)\, ,
\ee
where $k$ is used to label the level of the excitations. Accordingly, the eigen-energy associated with $\psi_{\rm aniso}^{[k]}({\bf r})$ is denoted as $E^{[k]}$.

With the {\em effective screening mass} proposed in Eq.~(\ref{effm0}), the anisotropic potential as given in Eq.~(\ref{kms}) can be expanded around an isotropic configuration $\tilde{m}_D (\lambda,\xi,\theta)= {\cal{M}}_{l^\prime m^\prime}(\lambda, \xi)$ with the leading order contribution is denoted by $V^{(0)}_{l^\prime m^\prime}(r) \equiv V(r, {\cal{M}}_{l^\prime m^\prime})$. The expansion can be formally expressed as the following Taylor series
\be\la{vex}
V({\bf r})\equiv V^{(0)}_{l^\prime m^\prime}(r) +\Delta V_{l^\prime m^\prime}({\bf r})=V^{(0)}_{l^\prime m^\prime}(r) +\sum_{s=1}^{\infty}V^{(s)}_{l^\prime m^\prime}({\bf r})\, ,
\ee
where the radial and angular dependences in the anisotropic terms can be separated as
\be\la{gf}
V^{(s)}_{l^\prime m^\prime}({\bf r})= {\cal{G}}^{(s)}_{l^\prime m^\prime} (\lambda,\xi,r)\cdot {\cal{F}}_{l^\prime m^\prime}^s(\lambda,\xi,\theta)\, ,
\ee
with
\be\la{defgf}
{\cal{G}}^{(s)}_{l^\prime m^\prime} (\lambda,\xi,r)\equiv \frac{{\cal{M}}_{l^\prime m^\prime}^s}{s!}\frac{\partial^s V({\bf r})}{\partial \tilde{m}^s_D}\bigg|_{\tilde{m}_D={\cal{M}}_{l^\prime m^\prime}}\, \quad\quad{\rm and}\quad \quad {\cal{F}}_{l^\prime m^\prime}(\xi,\theta)\equiv \frac{\tilde{m}_D(\lambda,\xi,\theta)}{{\cal{M}}_{l^\prime m^\prime}(\lambda,\xi)}-1\, .
\ee

If the Taylor series converges quickly, higher order terms $V_{l^\prime m^\prime}^{(s)}({\bf r})$ for $s\ge 1$ act as a small perturbation to the isotropic potential $V^{(0)}_{l^\prime m^\prime}(r)$. As a result, the eigen-energy $E_{nl}$ determined by Eq.~(\ref{raeq}) with $V(r)=V^{(0)}_{l^\prime m^\prime}(r)$ could be an ideal approximation to the eigen-energy $E^{[k]}$ of the bound state in an anisotropic medium. However, there exists an ambiguity due to fact that both the reduced radial wave-function $R_{nl}(r)$ and the eigen-energy $E_{nl}$ are not unique because the above isotropic potential constructed with the {\em effective screening mass} has a $(l^\prime , m^\prime)$- dependence. For example, with different values of $l^\prime$ and $m^\prime$, a set of ground states $\psi^{\rm iso}_{000}({\bf r})$ can be obtained based on Eqs.~(\ref{isowf}) and (\ref{raeq}). A question naturally arising is which values of $l^\prime$ and $m^\prime$ lead to an eigen-energy $E_{00}$ that is closest to $E^{[0]}$. This question can be answered by looking at the energy correction $\Delta E$ which in the first approximation in the perturbation theory reads
\be\la{ec00}
\Delta E\equiv \sum_{s=1}^{\infty} \Delta E_s= \sum_{s=1}^{\infty} \langle {\rm{Y}}_{00}(\theta,\phi)| {\cal{F}}_{l^\prime m^\prime}^s(\xi,\theta) | {\rm{Y}}_{00}(\theta,\phi)\rangle \cdot \langle R_{00}(r)| {\cal{G}}_{l^\prime m^\prime}^{(s)}(\lambda,\xi,r) | R_{00}(r)\rangle\, .
\ee
Naively, one can expect that the main contribution to the energy correction comes from the first term $\Delta E_1$. Therefore, when choosing $(l^\prime , m^\prime)=(0,0)$, this term vanishes by definition and the perturbative correction to the eigen-energy $E_{00}$ becomes negligible. The above conjecture indicates that the isotropic potential $V^{(0)}_{00}(r)$ constructed with the {\em effective screening mass} ${\cal {M}}_{00}$ is the right one that should be used in Eq.~(\ref{raeq}). The resulting ground state described by the wave-function $\psi^{\rm iso}_{000}({\bf r})$ is the ``most similar state"  in the sense that the associated eigen-energy $E_{00}$ is the best approximation to the lowest energy $E^{[0]}$ in an anisotropic medium. Furthermore, using the ``most similar state" is actually consistent with the explanation for the polarization of states with non-zero angular momentum  appearing in an anisotropic plasma which has already been discussed in Sec.~\ref{model}. 
This conclusion can be generalized to excited states. Solving the stationary SE with an isotropic potential $V^{(0)}_{l^\prime m^\prime}(r)$, among the eigen-states as formally given by Eq.~(\ref{isowf}), we are interested in a set of the states with the quantum numbers $(l, m)=(l^\prime, m^\prime)$. The eigen-energy $E^{[k]}$ can be well approximated by $E_{nl^\prime}$ of the so-called ``most similar state"  $\psi^{\rm iso}_{nl^\prime m^\prime}({\bf r})$.

In Table~\ref{corr00}, we list the eigen-energies as well as the perturbative corrections evaluated with different ${\cal {M}}_{l^\prime m^\prime}$ for the bottomonium states (including $\Upsilon$, $\chi_{b0}$ and $\chi_{b\pm1}$) at $\xi=1$ and $\lambda=1.1T_c$. Comparing with the exact values, $E^{[0]}=115.358\ {\rm MeV}$, $E^{[1]}=506.318\ {\rm MeV}$ and $E^{[2]}=507.878\ {\rm MeV}$ which are obtained by solving the three-dimensional SE with the anisotropic HQ potential in Eq.~(\ref{kms}), we find that the energies $E_{00}$ or $E_{01}$ associated with the ``most similar state" are closest to exact results as expected. The above discussion shows that one-dimensional SE should be sufficient to provide the desired information on the bound states in an anisotropic QCD medium at a quantitative level, namely,  the solution with an isotropic potential $V^{(0)}_{l^\prime m^\prime}(r)$ is indeed a very good approximation to that of an anisotropic medium characterized by a screening mass $\tilde{m}_D(\lambda,\xi,\theta)$.  In addition, finding excited states in a numerical approach is also a challenging task especially when the anisotropic situation is taken into account. On the other hand, when solving the Schr\"odinger equation with the $(l^\prime , m^\prime)$-dependent potential $V^{(0)}_{l^\prime m^\prime}(r)$ , the obtained ``ground state" $\psi_{0l^\prime m^\prime}({\bf r})$ for $(l^\prime , m^\prime)\neq(0,0)$ actually corresponds an excited state in the anisotropic medium. For example, the first two excited states of bottomonia denoted as $\chi_{b_{0}}$ and $\chi_{b_{\pm 1}}$ are identified with the ``ground states" determined by $V^{(0)}_{10}(r)$ and $V^{(0)}_{1\pm1}(r)$, respectively.  Therefore, finding the excited states can be also simplified in our approach based on the {\em effective screening mass}.

\begin{center}
\begin{table}[htpb]
\setlength{\tabcolsep}{7mm}{
\begin{tabular}{  c  c  c  c c  c}
\toprule[1pt]
$    \Upsilon       $ & $E_{00}$ & $\Delta E_1$ & $\Delta E_2$ & $     \Delta E_3      $  & $      \Delta E_4     $    \\ \hline
$ {\cal {M}}_{00}$ & $115.31$  & $ 0             $ & $ 0.013       $ & $1.5\times 10^{-4} $  & $9.2\times 10^{-6} $    \\
${\cal {M}}_{10}$  & $114.06$  & $1.25         $ & $ 0.026       $ & $-1.2\times 10^{-3}$  & $3.1\times 10^{-5} $    \\
${\cal {M}}_{11}$  & $115.96$  & $-0.64        $ & $ 0.015       $ & $6.7\times 10^{-4} $  & $1.9\times 10^{-5} $    \\ 	
${\cal {M}}_{20}$  & $114.53$  & $0.78         $ & $ 0.018       $ & $-5.6\times 10^{-4}$  & $1.5\times 10^{-5} $    \\
${\cal {M}}_{22}$  & $116.25$  & $-0.93        $ & $0.018        $ & $9.5\times 10^{-4} $  & $2.7\times 10^{-5} $    \\
\bottomrule[1pt]	
\end{tabular}
\vspace{0.2cm}
\vspace{0.2cm}
\begin{tabular}{  c  c  c  c c  c}
\toprule[1pt]
$  \chi_{b\pm1}  $ & $E_{01}$ & $\Delta E_1$ & $\Delta E_2$ & $     \Delta E_3      $   & $      \Delta E_4     $    \\ \hline
$ {\cal {M}}_{11}$ & $506.39$  & $ 0             $ & $ -0.005      $ & $-1.4\times 10^{-4}$  & $-5.7\times 10^{-6} $    \\
${\cal {M}}_{00}$  & $506.94$  & $-0.54        $ & $ -0.007      $ & $1.4\times 10^{-4} $  & $-5.4\times 10^{-6} $    \\
${\cal {M}}_{10}$  & $507.97$  & $-1.48        $ & $ -0.023      $ & $1.3\times 10^{-3} $  & $-4.1\times 10^{-5} $    \\ 	
${\cal {M}}_{20}$  & $507.59$  & $-1.14        $ & $ -0.015      $ & $7.3\times 10^{-4} $  & $-2.0\times 10^{-5} $    \\
${\cal {M}}_{22}$  & $506.13$  & $0.26         $ & $ -0.005      $ & $-2.5\times 10^{-4}$  & $-8.7\times 10^{-6} $    \\
\bottomrule[1pt]	
\end{tabular}
\vspace{0.2cm}
\vspace{0.2cm}
\begin{tabular}{  c  c  c  c c  c}
\toprule[1pt]
$  \chi_{b0}        $ & $E_{01}$ & $\Delta E_1$ & $\Delta E_2  $ & $     \Delta E_3      $  & $      \Delta E_4     $    \\ \hline
$ {\cal {M}}_{10}$ & $507.97$  & $ 0             $ & $ -0.006       $ & $9.6\times 10^{-5} $  & $-4.9\times 10^{-6} $    \\
${\cal {M}}_{00}$  & $506.94$  & $1.07         $ & $ -0.012       $ & $-6.6\times 10^{-4}$  & $-1.9\times 10^{-5} $    \\
${\cal {M}}_{11}$  & $506.39$  & $1.67         $ & $ -0.020       $ & $-1.5\times 10^{-3}$  & $-5.2\times 10^{-5} $    \\ 	
${\cal {M}}_{20}$  & $507.59$  & $0.38         $ & $ -0.007       $ & $-1.1\times 10^{-4}$  & $-5.2\times 10^{-6} $    \\
${\cal {M}}_{22}$  & $506.13$  & $1.97         $ & $-0.024        $ & $-2.0\times 10^{-3}$  & $-7.8\times 10^{-5} $    \\
\bottomrule[1pt]	
\end{tabular}
}
\caption{The eigen-energies and their perturbative corrections evaluated with different  {\em effective screening masses} for the bottomonium states $\Upsilon$, $\chi_{b\pm1}$ and $\chi_{b0}$. The results are obtained at $\xi=1$, $\lambda=1.1T_c$ and given in the unit of {\rm {MeV}}.}
\label{corr00}
\end{table}
\end{center}

Furthermore, the Taylor series given in Eq.~(\ref{vex}) is very different from the expansion of the anisotropic potential $V({\bf r})$ around $\xi=0$. Although the leading contributions are both independent of the angles, the energy splitting of states with angular quantum number $l\neq 0$ already shows up at leading order due to the $m$-dependence of the {\em effective screening mass} when Eq.~(\ref{vex}) is considered. On the other hand, in order to see such a splitting of energy in the latter approach, one has to take into account the anisotropic higher order corrections. Of course, the main disadvantage of the expansion around $\xi=0$ is the limitation for application at large anisotropies.

\section{discussion on the suppression of the Taylor series}
\la{suppression}

The ``most similar state"  $\psi^{\rm iso}_{n l^\prime m^\prime}({\bf r})$ is an eigen-state determined by Eqs.~(\ref{isowf}) and (\ref{raeq}) where an isotropic heavy-quark potential $V(r,{\cal {M}}_{l^\prime m^\prime})$ should be taken into account.   As shown in Table~\ref{corr00}, the perturbative corrections to the eigen-energy are so small that $\psi^{\rm iso}_{n l^\prime m^\prime}({\bf r})$ could be an ideal approximation to the corresponding eigen-state $\psi_{\rm aniso}^{[k]}({\bf r})$ in an anisotropic plasma. Therefore, the one-dimensional SE with the {\em effective screening mass} can be used to study quarkonia in an anisotropic medium. For any given quantum numbers $n$, $l$ and $m$, the eigen-energy corrections can be written as
\be\la{ec}
\Delta E\equiv \sum_{s=2}^{\infty} \Delta E_s= \sum_{s=2}^{\infty} \langle {\rm{Y}}_{lm}(\theta,\phi)| {\cal{F}}_{l m}^s(\xi,\theta) | {\rm{Y}}_{lm}(\theta,\phi)\rangle \cdot \langle R_{nl}(r)| {\cal{G}}_{l m}^{(s)}(\lambda,\xi,r) | R_{nl}(r)\rangle\, .
\ee
Notice that different from Eq.~(\ref{ec00}), the sum in the above equation starts from $s=2$ because only the ``most similar state" will be considered in the following. In order to understand the reason why the contribution from Eq.~(\ref{ec}) is negligible, we formaly divided $\Delta E$ into the following two parts
\be
\Delta E\equiv \sum_{s=2}^{\infty} \Delta E_s= \sum_{s=2}^{\infty}  \Delta E^{{\rm ag}}_s\cdot \Delta E^{{\rm ra}}_s\, ,
\ee
where
\be
\la{eag}
\Delta E^{{\rm ag}}_s =  \langle {\rm{Y}}_{lm}(\theta,\phi)| {\cal{F}}_{l m}^s(\xi,\theta) | {\rm{Y}}_{lm}(\theta,\phi)\rangle \, ,
\ee
and
\be
\la{era}
\Delta E^{{\rm ra}}_s = \langle R_{nl}(r)| {\cal{G}}_{l m}^{(s)}(\lambda,\xi,r) | R_{nl}(r)\rangle \, .
\ee

\subsection{The angular part of the perturbative corrections to the eigen-energies}

We first look at the angular part. Since the spherical harmonics are universal, Eq.~(\ref{eag}) can be determined once the exact form of the anisotropic Debye mass is known. This means $\Delta E^{{\rm ag}}_s$ is independent of the specific potential model used in the Schr\"odinger equation. Starting with the case where $\xi$ is small, we have
\be
\Delta E^{{\rm ag}}_s(\xi \ll 1) = (\frac{\xi}{8} )^s\langle {\rm{Y}}_{lm}(\theta,\phi)| (x^2 +1-k(l,m))^s | {\rm{Y}}_{lm}(\theta,\phi)\rangle \, , \label{eq22}
\ee
where $x\equiv \cos \theta$. One can expect that with increasing $s$, the (absolute) values of $\Delta E^{{\rm ag}}_s(\xi \ll 1)$ get decreased simply due to the factor $(\xi/8)^s$. In fact, we can further show that there is an extra suppression from the $(l,m)-$dependent term in the above equation which is denoted as
\ba
{\cal I}_{lm}^{(s)} &\equiv & \langle {\rm{Y}}_{lm}(\theta,\phi)| (x^2 +1-k(l,m))^s | {\rm{Y}}_{lm}(\theta,\phi)\rangle\, \nonumber \\
&=& \sum_{n=0}^{s} {\rm C}_s^n (1-k(l,m))^{s-n} {\cal{X}}^{(2n)}\, .
\ea
Here, ${\rm C}_s^n$ is the binomial coefficient and ${\cal{X}}^{(n)}$ is defined as ${\cal{X}}^{(n)}=\langle {\rm{Y}}_{lm}| x^n | {\rm{Y}}_{lm}\rangle$. Using the iterative formula for the spherical harmonics
\be\la{ir}
x {\rm{Y}}_{lm}=a_{l,m} {\rm{Y}}_{l+1,m}+a_{l-1,m} {\rm{Y}}_{l-1,m} \quad\quad {\rm with} \quad \quad a_{lm}=\sqrt{\frac{(l+1)^2-m^2}{(2l+1)(2l+3)}}\, ,
\ee
it is straightforward to obtain an analytical expression for ${\cal{X}}^{(n)}$, although this is rather tedious for large $n$. Explicitly, for $s=2$, we arrive at
\be\la{i2}
{\cal I}_{lm}^{(2)}={\cal{X}}^{(4)}+2(1-k(l,m)){\cal{X}}^{(2)}+(1-k(l,m))^2 \, ,
\ee
with
\be
{\cal{X}}^{(2)}= a_{lm}^2+a_{l-1,m}^2\quad\quad{\rm and}\quad\quad {\cal{X}}^{(4)}= a_{lm}^2(a_{l+1,m}^2+a_{lm}^2+a_{l-1,m}^2)+a_{l-1,m}^2(a_{lm}^2+a_{l-1,m}^2+a_{l-2,m}^2)\,.\label{eq26}
\ee
In fact, one can prove that Eq.~(\ref{i2}) approaches zero when $l\rightarrow \infty$ and $m=\pm l$ and its maximum value $68/441$ can be obtained when $l=2$ and $m=0$. Therefore, ${\cal I}_{lm}^{(2)} \le 68/441\approx0.154$.

For large $s$, the $(l,m)-$dependence in ${\cal I}_{lm}^{(s)}$ turns to be very complicated. Instead, we will consider the following ratio for $n=1,2,3,\cdots$
\be
\frac{|{\cal I}_{lm}^{(2n+1)}|}{{\cal I}_{lm}^{(2n)}}<\frac{ \langle {\rm{Y}}_{lm}(\theta,\phi)| (x^2 +1-k(l,m))^{2n} |x^2 +1-k(l,m)| | {\rm{Y}}_{lm}(\theta,\phi)\rangle}{\langle {\rm{Y}}_{lm}(\theta,\phi)| (x^2 +1-k(l,m))^{2n} | {\rm{Y}}_{lm}(\theta,\phi)\rangle} < \delta_{lm} \, .
\ee
Here, $\delta_{lm}$ is defined as the maximum value of $ |x^2 +1-k(l,m)|$ for a given $(l,m)$ when $x^2$ changes from $0$ to $1$. Since $1\le k(l,m) \le 8/5$, $\delta_{lm}$ can not be larger than $1$. In fact, the largest $\delta_{lm}$ is given by $\delta_{\infty, \pm l}=1$.
In addition, ${\cal I}_{lm}^{(2n+2)}/{\cal I}_{lm}^{(2n)}<\delta_{lm}^2$ can be also shown similarly.

Based on the above discussion, we conclude that the magnitude of ${\cal I}_{lm}^{(s)}$ is small and gets suppressed when $s$ is large. In particular, the maximum value of ${\cal I}_{lm}^{(2)}$ is given by $0.154$. For $s>2$, although the exact values of the maximum are not computed here, it is possible to estimate an upper limit which is given by $|{\cal I}_{lm}^{(s)}|< {\cal I}_{lm}^{(2)} \delta_{lm}^{s-2}(l,m) $. Given the quantum numbers $l$ and $m$, explicitly we have
\be
|{\cal I}_{lm}^{(s>2)}|<\begin{cases}
\frac{4}{45}(\frac{2}{3})^{s-2} & {\rm for} \quad l=0,m=0;\\
\frac{12}{175}(\frac{3}{5})^{s-2} & {\rm for} \quad l=1,m=0;\\
\frac{8}{175}(\frac{4}{5})^{s-2} &{\rm for} \quad l=1,m=\pm1;\\
\frac{68}{441}(\frac{11}{21})^{s-2} & {\rm for} \quad l=2,m=0.
\end{cases}
\ee

We should mention that for even $s$, ${\cal I}_{lm}^{(s)}$ is monotonically decreasing with increasing $s$, however, the same conclusion can not be drawn for $s=2,3,4,\cdots$. Instead, only ${\cal I}_{lm}^{(2n)}>|{\cal I}_{lm}^{(2n+1)}|$ for $n=1,2,3,\cdots$ can be justified. Furthermore, the iterative formula in Eq.~(\ref{ir}) turns to be very useful to analytically study the angular part of the perturbative energy correction even for the case of arbitrary anisotropies as long as the anisotropic Debye mass can be parameterized as a polynomial function of $x$, namely $\tilde{m}_D(\lambda,\xi,\theta)=\sum_{i=0}^{n} c_i(\lambda,\xi) x^i$. However, $\tilde{m}_D(\lambda,\xi,\theta)$ as given in Eq.~(\ref{eff11}) doesn't satisfy this condition.

\begin{figure}[htbp]
\begin{center}
\includegraphics[width=0.5\textwidth]{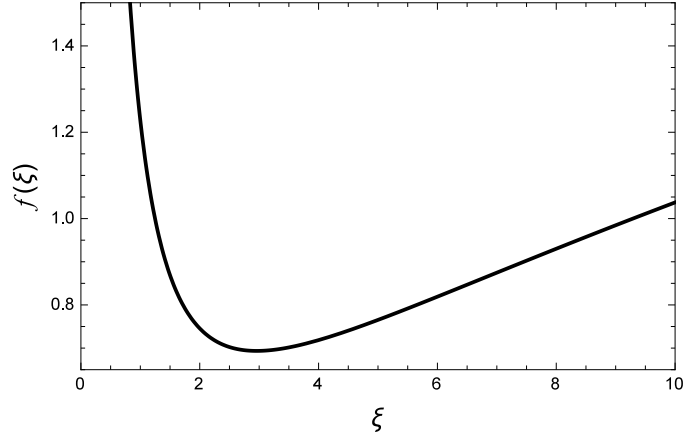}
\caption{The $\xi$-dependence of $f(\xi)$.}
\label{fx}
\end{center}
\end{figure}

For arbitrary $\xi$, we start with the simplest case where $s=2$
\ba\la{ag2g}
\Delta E^{{\rm ag}}_2(\xi) &=& \frac{\langle {\rm{Y}}_{lm}(\theta,\phi)| \tilde{m}^2_D(\lambda,\xi,\theta)| {\rm{Y}}_{lm}(\theta,\phi)\rangle }{{\cal{M}}^2_{l m}(\lambda,\xi)}-1 \,  \nonumber\\
&=& \frac{\langle {\rm{Y}}_{lm}(\theta,\phi)| (x^2+f(\xi))^{-1/2}| {\rm{Y}}_{lm}(\theta,\phi)\rangle }{\Big(\langle {\rm{Y}}_{lm}(\theta,\phi)|(x^2+f(\xi))^{-1/4}| {\rm{Y}}_{lm}(\theta,\phi)\rangle\Big)^2}-1\, .
\ea
In the above equation $f(\xi)=f_2(\xi)/f_1(\xi)$ and the explicit expression of $f(\xi)$ is rather complicated which we don't list here. However, the key point is that $f(\xi)$ is positive for $\xi>0$ and it is not a monotonic function of $\xi$.  The minimum can be numerically determined as $f_{\rm min}(\xi)\approx f(2.958)\approx 0.694$. In Fig.~\ref{fx}, we plot $f(\xi)$ as a function of $\xi$.

\begin{figure}[ht]
\begin{center}
\vspace{2mm}
\includegraphics[width=0.46\textwidth]{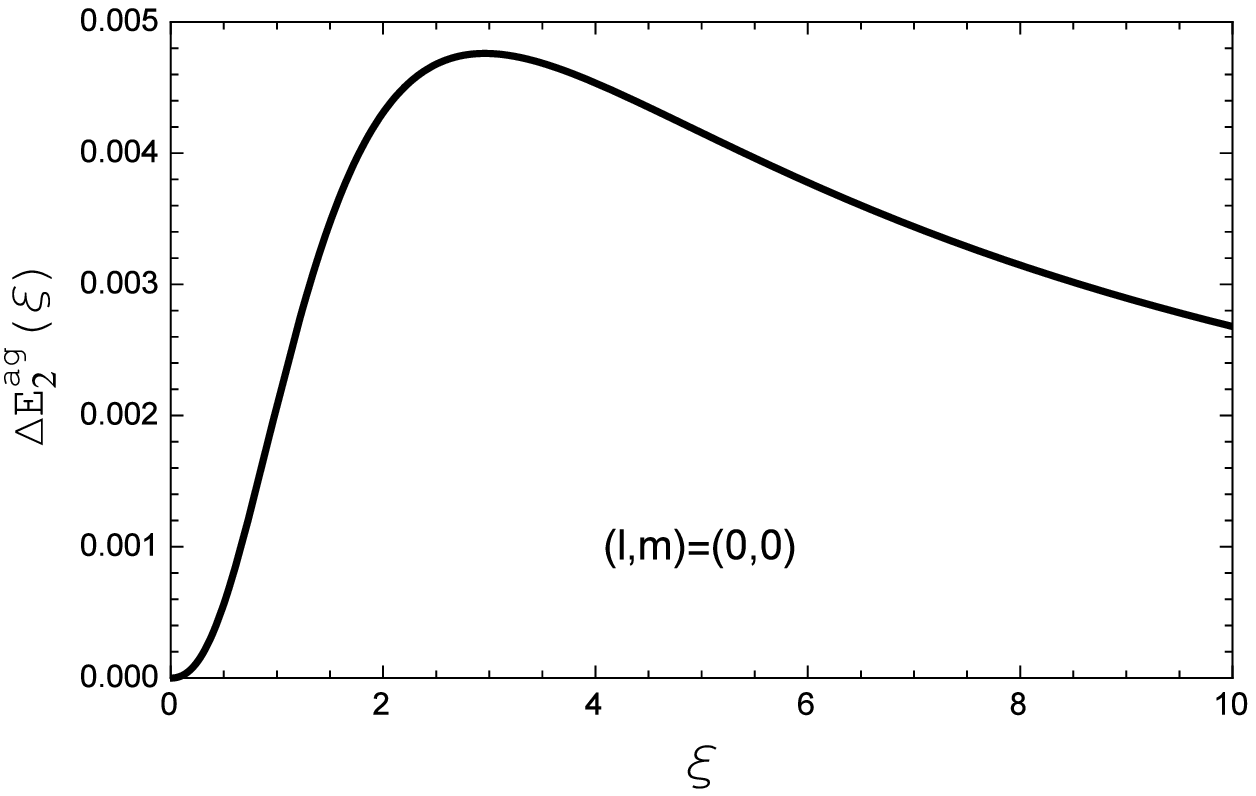} \hspace{5mm}
\includegraphics[width=0.46\textwidth]{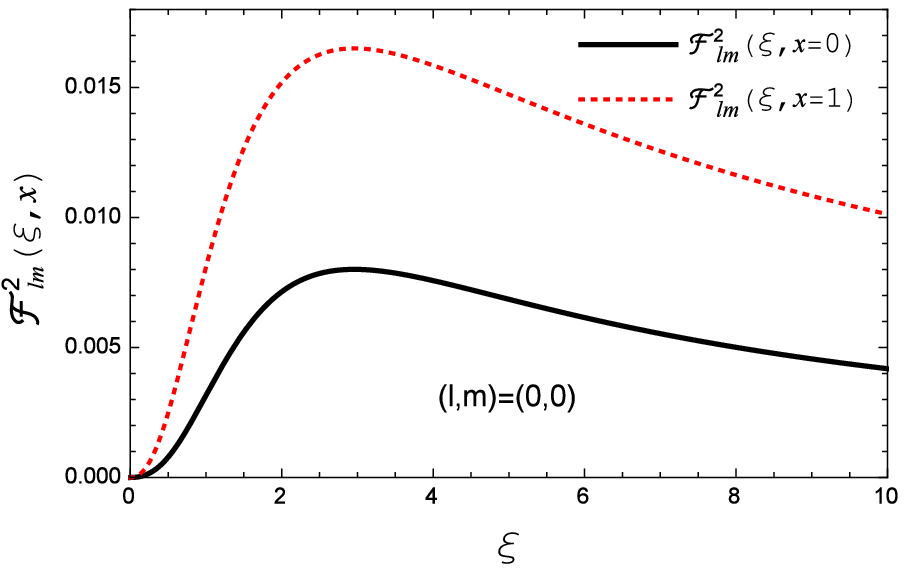}
\end{center}
\vspace{-4mm}
\caption{Left panel shows the $\xi$-dependence of $\Delta E^{{\rm ag}}_2(\xi)$ at $l=0$ and $m=0$. Right panel shows the $\xi$-dependence of $ {\cal{F}}_{l m}^{2}(\xi, x=0)$ and $ {\cal{F}}_{l m}^{2}(\xi, x=1)$ at $l=0$ and $m=0$.}
\label{fig:dE2agF2}
\end{figure}

It is not easy to get a general expression of $\Delta E^{{\rm ag}}_2(\xi) $ for any given $l$ and $m$. On the other hand, in order to see the non-trivial $\xi$-dependence of $\Delta E^{{\rm ag}}_2(\xi) $, we fix $l$ and $m$. The simplest case is to consider $(l,m)=(0,0)$ where the numerator in Eq.~(\ref{ag2g}) can be carried out analytically and the result takes the following form
\be
\Delta E^{{\rm ag}}_2(\xi)=\frac{\sqrt{f(\xi)} \ln \Big(\frac{\sqrt{1+f(\xi)}+1}{\sqrt{f(\xi)}}\Big)}{\Big(\int _0^1 (1+x^2/f(\xi))^{-1/4} {\rm d} x\Big)^2}-1\, ,\quad\quad{\rm with}\quad\quad l=0,\,m=0\, ,\label{eq30}
\ee
where the integral in the denominator corresponds to a hypergeometric function. Numerically, it can be shown that the above result decreases with increasing values of $f(\xi)$, therefore, the maximum of $\Delta E^{{\rm ag}}_2(\xi)$ with $(l,m)=(0,0)$ appears when $\xi\approx 2.958$, see Fig.~\ref{fig:dE2agF2} (the left panel). Changing the values of $l$ and $m$, we have checked that the maximum of $\Delta E^{{\rm ag}}_2(\xi)$ is always at $\xi\approx 2.958$ and the maximum values of $\Delta E^{{\rm ag}}_2(\xi)$ at a given $(l,m)$ are list in Table~\ref{ag2}. Our numerical results also suggest that as a function of $l$ and $m$, $\Delta E^{{\rm ag}}_2(2.958)$ becomes largest when $(l,m)=(2,0)$ which coincides with the finding in the small $\xi$ limit.

\begin{center}
\begin{table}[htpb]
\setlength{\tabcolsep}{4mm}{
\begin{tabular}{ |c|  c|  c|  c| c|  c|}
\hline
     \backslashbox{$m$}{$l$} & $     0    $  &  $    1   $  &  $     2    $  &  $    3     $  &  $    10    $    \\ \hline
	  0  & $ 0.00476 $    &    $0.00307$  &    $ 0.00823 $  &    $ 0.00600 $  &  $ 0.00621 $    \\ \hline
	  1  & $\backslash$  &    $0.00281$  &    $ 0.00279 $  &    $ 0.00605 $  &  $ 0.00614 $    \\ \hline
	  2  & $\backslash$  &  $\backslash$&    $ 0.00186 $  &    $ 0.00230 $  &  $ 0.00591 $    \\ \hline	
	  3  & $\backslash$  &  $\backslash$&  $\backslash$  &    $ 0.00132 $  &  $ 0.00553 $    \\ \hline
	 10  & $\backslash$  &  $\backslash$&  $\backslash$  &  $\backslash$  &  $ 0.00032 $    \\ \hline	
\end{tabular}
}
\caption{The maximum values of $\Delta E^{{\rm ag}}_2(\xi)$ at different $l$ and $m$.}
\label{ag2}
\end{table}
\end{center}

For $s>2$, we are not going to evaluate $\Delta E^{{\rm ag}}_s(\xi)$ with various quantum numbers $(l,m)$, instead, we try to estimate the corresponding upper limit which should be sufficient to demonstrate the suppression of $\Delta E^{{\rm ag}}_s(\xi)$ when $s$ is large. It is clear that the function
\ba
 {\cal{F}}_{l m}(\xi,x) = \frac{(x^2+f(\xi))^{-1/4} }{\langle {\rm{Y}}_{lm}(\theta,\phi)|(x^2+f(\xi))^{-1/4}| {\rm{Y}}_{lm}(\theta,\phi)\rangle}-1\, ,
\ea
is monotonically decreasing with $x^2$ when $x$ changes from $0$ to $1$, therefore, the maximum of ${\cal{F}}^2_{l m}(\xi,x)$ locates at either $x=0$ or $x=1$ which can be denoted as $max ( {\cal{F}}_{l m}^{2}(\xi, x=0),{\cal{F}}_{l m}^{2}(\xi, x=1))$. Therefore, the ratio of $\Delta E^{{\rm ag}}_{2n+2}(\xi)/\Delta E^{{\rm ag}}_{2n}(\xi)$ for $n=1,2,3,\cdots$, satisfies
\be
\frac{\Delta E^{{\rm ag}}_{2n+2}(\xi)}{\Delta E^{{\rm ag}}_{2n}(\xi)}=\frac{ \langle {\rm{Y}}_{lm}(\theta,\phi)|  {\cal{F}}_{l m}^{2n+2}(\xi,\theta)| {\rm{Y}}_{lm}(\theta,\phi)\rangle}{\langle {\rm{Y}}_{lm}(\theta,\phi)|  {\cal{F}}_{l m}^{2n}(\xi,\theta) | {\rm{Y}}_{lm}(\theta,\phi)\rangle}< max( {\cal{F}}_{l m}^{2}(\xi, x=0),{\cal{F}}_{l m}^{2}(\xi, x=1))\, .
\ee
Then, the question amounts to estimating  an upper limit for $ {\cal{F}}_{l m}^{2}(\xi, x=0)$ and ${\cal{F}}_{l m}^{2}(\xi, x=1)$ for arbitrary anisotropies and the quantum numbers. Notice that $f(\xi)>0$ and $f_{\rm min}(\xi)\approx 0.694$, we have
\ba
 {\cal{F}}_{l m}^{2}(\xi, x=0)&=&\Big(\frac{1}{\langle {\rm{Y}}_{lm}(\theta,\phi)|(f^{-1}(\xi)x^2+1)^{-1/4}| {\rm{Y}}_{lm}(\theta,\phi)\rangle}-1\Big)^2<\Big(\frac{1}{(1/f(\xi)+1)^{-1/4}}-1\Big)^2\, \nonumber \\
& \le & \Big(\frac{1}{(1/f_{\rm min}(\xi)+1)^{-1/4}}-1\Big)^2\approx 0.0625\, .
 \ea
The estimated upper limit is valid independent on the values of $(l,m)$. The analysis on $ {\cal{F}}_{l m}^{2}(\xi, x=1)$ can be carried out in a similar approach and we can show
\ba
 {\cal{F}}_{l m}^{2}(\xi, x=1)&=&\Big(\frac{1}{\langle {\rm{Y}}_{lm}(\theta,\phi)|\big(\frac{x^2}{1+f(\xi)}+\frac{f(\xi)}{1+f(\xi)}\big)^{-1/4}| {\rm{Y}}_{lm}(\theta,\phi)\rangle}-1\Big)^2<\Big(\frac{1}{(f(\xi)/(f(\xi)+1))^{-1/4}}-1\Big)^2\, \nonumber \\
& \le & \Big(\frac{1}{(f_{\rm min}(\xi)/(f_{\rm min}(\xi)+1))^{-1/4}}-1\Big)^2\approx 0.0400\, .
 \ea
On the other hand, once $l$ and $m$ are specified, ${\cal{F}}_{l m}^{2}(\xi, x=0)$ and $ {\cal{F}}_{l m}^{2}(\xi, x=1)$ can be evaluated numerically and one can further reduce this limit. As shown in Fig.~\ref{fig:dE2agF2} (the right panel), with $(l,m)=(0,0)$, the maxima of ${\cal{F}}_{l m}^{2}(\xi, x=0)$ and ${\cal{F}}_{l m}^{2}(\xi, x=1)$ which appear at $\xi=2.958$ are found to be $0.0080$ and $0.0165$, respectively. More results are given in Table~\ref{ag3}.

\begin{center}
\begin{table}[htpb]
\setlength{\tabcolsep}{2mm}{
\begin{tabular}{ |c|  c|  c|  c| c|  c| c|}
\hline
    $(l,m)$ & $     (0,0)    $  &  $    (1,0)   $  &  $     (1,\pm1)    $  &  $    (2,0)     $  &  $    (2,\pm1)    $ &  $    (2,\pm2)    $   \\ \hline
	   ${\cal{F}}_{l m}^{2}(\xi=2.958, x=0)$  & $ 0.0080 $    &    $0.0257$  &    $ 0.0033 $  &    $ 0.0167 $  &  $ 0.0144 $ &  $ 0.0018 $    \\ \hline
	   ${\cal{F}}_{l m}^{2}(\xi=2.958, x=1)$  & $0.0165$  &    $0.0051$  &    $ 0.0238 $  &    $ 0.0093 $  &  $ 0.0108 $ &  $ 0.0275 $      \\ \hline	
\end{tabular}
}
\caption{The maximum values of ${\cal{F}}_{l m}^{2}(\xi, x=0)$ and ${\cal{F}}_{l m}^{2}(\xi, x=1)$ at different $l$ and $m$.}
\label{ag3}
\end{table}
\end{center}

At this point, the estimation that $max( {\cal{F}}_{l m}^{2}(\xi, x=0),{\cal{F}}_{l m}^{2}(\xi, x=1))<0.0625$ is justified.  In addition, the square root of this upper limit corresponds to that of the ratio $|\Delta E^{{\rm ag}}_{2n+1}(\xi)|/\Delta E^{{\rm ag}}_{2n}(\xi)$, namely, the following relations are always true regardless of the specific values of $\xi$ as well as the quantum numbers $l$ and $m$,
\be
\Delta E^{{\rm ag}}_{2n+2}(\xi) < 0.0625 \Delta E^{{\rm ag}}_{2n}(\xi)\, ,\quad\quad{\rm and}\quad\quad |\Delta E^{{\rm ag}}_{2n+1}(\xi)| < 0.250 \Delta E^{{\rm ag}}_{2n}(\xi)\, .
\ee

It is similar as what we found for the small $\xi$ case that the above result doesn't guarantee the relation that $|\Delta E^{{\rm ag}}_{2n+3}(\xi)|<|\Delta E^{{\rm ag}}_{2n+1}(\xi)|$ or $|\Delta E^{{\rm ag}}_{2n+2}(\xi)|<|\Delta E^{{\rm ag}}_{2n+1}(\xi)|$. This is because the absolute values of $\Delta E^{{\rm ag}}_{2n+1}(\xi)$ could be very close to zero even for small $n$. One example we found is that with $(l,m)=(2,1)$, $\Delta E^{{\rm ag}}_{4}(\xi)>|\Delta E^{{\rm ag}}_{3}(\xi)|$ when $\xi$ is around $1$.

For practical applications, we are probably more interested in quarkonium states with not very large azimuthal quantum number. Based on the results given in Tables~\ref{ag2} and \ref{ag3}, we can show the following
\be\la{de2}
|\Delta E^{{\rm ag}}_{s\ge 2}(\xi)|<\begin{cases}
0.00476(0.128)^{s-2} & {\rm for} \quad l=0,m=0;\\
0.00307(0.160)^{s-2} & {\rm for} \quad l=1,m=0;\\
0.00281(0.154)^{s-2} & {\rm for} \quad l=1,m=\pm1;\\
0.00823(0.129)^{s-2} & {\rm for} \quad l=2,m=0,
\end{cases}
\ee
which clear indicates that the absolute values of $\Delta E^{{\rm ag}}_{s}(\xi)$ are very small and decrease quickly with increasing $s$.

\subsection{The radial part of the perturbative corrections to the eigen-energies}

The discussions above of the angular part of the perturbative corrections depend only on the parameterization of the anisotropic Debye mass $\tilde{m}_D$. On the other hand, one has to specify an explicit form of the heavy-quark potential in order to study the corresponding radial part $\Delta E^{{\rm ra}}_s$ as defined in Eq.~({\ref{era}}). In addition, unlike the spherical harmonics $Y_{lm}(\theta,\phi)$, the radial wave-function $R_{nl}(r)$ can, in general, only be obtained numerically by solving Eq.~(\ref{raeq}) .

\begin{figure}[htbp]
\centering
\includegraphics[width=0.45\textwidth]{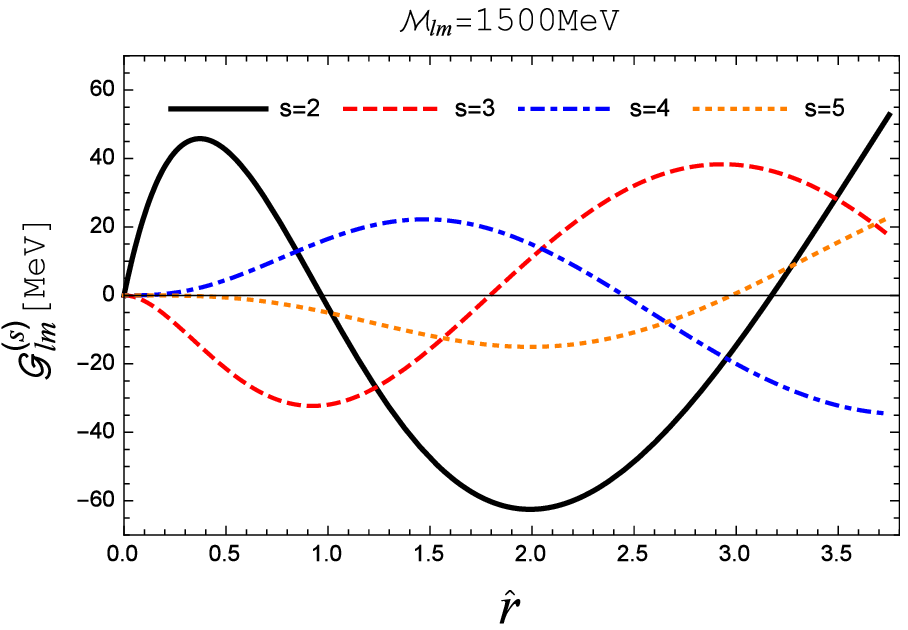}
\includegraphics[width=0.45\textwidth]{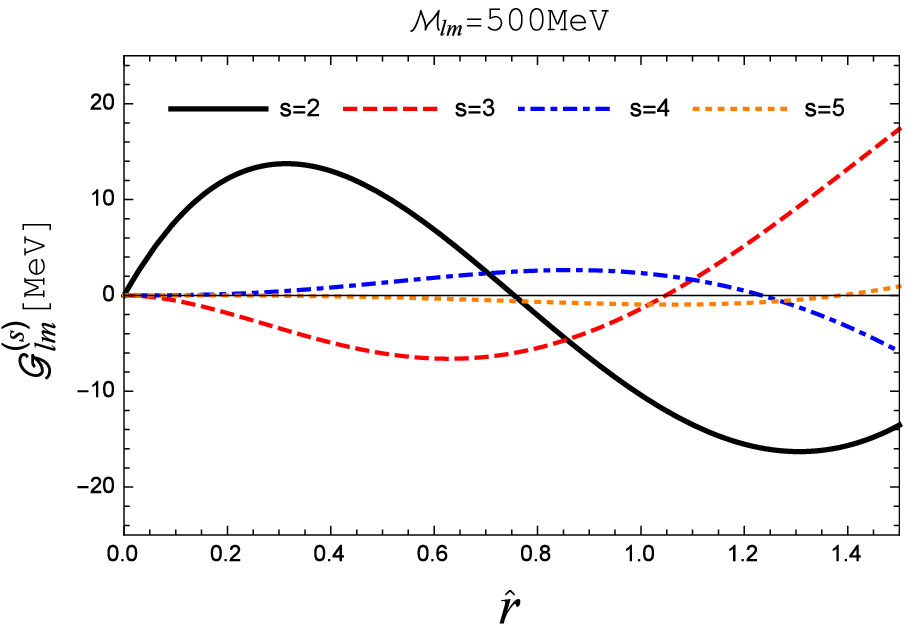}
\caption{${\cal{G}}^{(s)}_{lm}$ as a function of the dimensionless variable $\hat{r}\equiv r {\cal{M}}_{lm}$ for $s=2,3,4,5$. Left: The  {\em effective screening mass} ${\cal{M}}_{lm}=1500\ {\rm MeV}$. Right: the  {\em effective screening mass} ${\cal{M}}_{lm}=500\ {\rm MeV}$. }
\label{gs}
\end{figure}

To proceed further, we consider Eq.~(\ref{kms}) as the heavy-quark potential model. The explicit form of ${\cal{G}}^{(s)}_{lm}$ defined in Eq.~({\ref{defgf}}) can be written as
\be
{\cal{G}}^{(s)}_{lm} = \alpha {\cal{M}}_{lm} (-1)^{s-1} \frac{\hat{r}-(s-1)}{s!} \hat{r}^{s-1} e^{-\hat{r}} + \frac{2 \sigma}{{\cal{M}}_{lm}} (-1)^{s-1}\Big(\sum_{n=0}^{s}\frac{\hat{r}^{n}}{n!}+\frac{\hat{r}^{s+1}}{2s!}-e^{\hat{r}}\Big) e^{-\hat{r}}\, .\label{eq37}
\ee
To obtain the above equation, we have used the following derivatives
\be
\frac{d^s}{d x^s}\left(x e^{- r x}\right) = s (-r)^{s-1}e^{-r x}+x (-r)^s e^{-r x} \quad \mathrm{and}\quad \frac{d^s}{d x^s}\left(e^{- r x}/x\right) =\sum_{n=0}^{s} (-1)^s \frac{s!}{n!} \frac{r^{n}}{x^{s+1-n}} e^{-r x}\, .
\ee
In Fig.~\ref{gs}, we show ${\cal{G}}^{(s)}_{lm}$ as a function of $\hat{r}\equiv r {\cal{M}}_{lm}$ for $s=2,3,4,5$. The plot on the left corresponds to the  {\em effective screening mass} ${\cal{M}}_{lm}=1500\ {\rm MeV}$ and we focus on a region of the dimensionless variable up to $\hat{r}=3.75$. Therefore, the size of the quarkonium states are assumed to be smaller than $0.5\ {\rm fm}$, which roughly speaking is the upper limit of the root-mean-square radii of quarkonia above the critical temperature.  The right plot shows the results at ${\cal{M}}_{lm}=500\ {\rm MeV}$. Correspondingly, we consider a relatively narrow region of $\hat{r}$.

For the ``most similar state", due to the vanishing angular part in the perturbative correction, $\Delta E_1=0$ which is independent of the values of ${\cal{G}}^{(1)}_{lm}$. On the other hand, the plots in Fig.~\ref{gs} suggest that the magnitude of ${\cal{G}}^{(2)}_{lm}$ doesn't exceed $\sim 70\ {\rm MeV}$ within the given region of $\hat{r}$ that is relevant to quarkonium studies\footnote{The magnitude of ${\cal{G}}^{(s)}_{lm}$ increases if a wider region of $\hat{r}$ is considered. It eventually approaches to $2\sig/{\cal{M}}_{lm}$ when $\hat{r}\rightarrow \infty$.}. Therefore, $\Delta E_2$ is estimated to be less than $0.6\ {\rm MeV}$ given that $\Delta E^{{\rm ag}}_2\le 0.00823$, see Table~\ref{ag2}. In the same region of $\hat{r}$, the magnitude of ${\cal{G}}^{(s)}_{lm}$ gets smaller as $s$ is increasing, naively, we don't expect any enhancement from the radial part to $\Delta E_s$ when $s$ is large.

As a very rough estimation, the above discussion presumes that $R_{nl}(r)$ is simply a Dirac delta and becomes non-vanishing only if $r$ equals the root-mean-square radius of the bound state. To do it in a relatively rigorous way, we consider a normalized function
\be
|\psi(r)\rangle =2 \big(\frac{1}{a_0}\big)^{3/2} e^{-r/a_0}\, .
\ee
This is actually the radial part of the wave-function for ground state when the potential is Coulombic. However, the Bohr radius $a_0$ should be understood as the most probable radius of a quarkonium state. For the Coulomb term, one obtains
\ba\la{dec}
\Delta E^{{\rm ra, C}}_s &=& \alpha {\cal{M}}_{lm} (-1)^{s-1}  \langle R_{nl}(r)| \frac{\hat{r}-(s-1)}{s!} \hat{r}^{s-1} e^{-\hat{r}}  | R_{nl}(r)\rangle\, \nonumber\\
& \approx &  \alpha {\cal{M}}_{lm} (-1)^{s-1} \int_0^\infty  \frac{4 r^2}{a_0^3} e^{-2 r/a_0}  \frac{\hat{r}-(s-1)}{s!} \hat{r}^{s-1} e^{-\hat{r}}  d r \, \nonumber\\
&=& 4\alpha {\cal{M}}_{lm} (s+1) (2-2s+3{\hat a}_0)\frac{(-{\hat a}_0)^{s-1}}{(2+{\hat a}_0)^{s+3}}\, ,
\ea
with ${\hat a}_0\equiv {\cal{M}}_{lm} a_0$. Here, we have used the following integral
\be\label{in1}
\int_0^\infty e^{-a x} x^{n} d x =\frac{n!}{a^{n+1}}\,  ,
\ee
where $a>0$ and $n$ is a positive integer. Similarly, the contribution from the string term is given by
\be\la{des}
\Delta E^{{\rm ra, S}}_s = 2 \sig a_0 [2(s+3)(s+2)(1-s)+2{\hat a}_0(s+3)(s+4)+2{\hat a}_0^2(s+4)+{\hat a}_0^3]\frac{(-{\hat a}_0)^{s}}{(2+{\hat a}_0)^{s+4}}\, .
\ee
Notice that after integrating over $r$ with the help of Eq.~(\ref{in1}), the summation over $n$ appearing in Eq.~(\ref{eq37}) can be carried out by using the identity
\be
\sum_{n=0}^{s} \frac{(n+1)(n+2)}{(1+x)^{n+3}}=\frac{2}{x^3}-\frac{(s+3)(s+2)x^2+2(s+3)x+2}{(x+1)^{s+3}x^3}\, .
\ee

Both $\Delta E^{{\rm ra, C}}_s$ and $\Delta E^{{\rm ra, S}}_s$ have non-trivial dependences on the most probable radius $a_0$ as well as the {\em effective screening mass} ${\cal{M}}_{lm}$. In order to estimate an upper bound for the the radial part of the perturbative corrections $|\Delta E^{{\rm ra}}_s|\equiv |\Delta E^{{\rm ra, C}}_s+\Delta E^{{\rm ra, S}}_s|$, the value ranges of $a_0$ and ${\cal{M}}_{lm}$ have to be specified. When the screening effect becomes strong enough, even the smallest quarkonium state $\Upsilon$ can not survive. Therefore, it is reasonable to assume ${\cal{M}}_{lm} \le 1500\ {\rm {MeV}}$ which, based on the potential model adopted here, corresponds to a temperature less than $3T_c$ in the limit $\xi=0$. On the other hand, the most probable radius $a_0$ equals $\sqrt{\langle r^2 \rangle} /\sqrt{3}$, as a result, $a_0=0.3\ {\rm fm}$ is equivalent to a root-mean-square radius $\sqrt{\langle r^2 \rangle} \approx 0.5\ {\rm fm}$ which has been considered as the largest size of quarkonia that could be bound in the deconfining phase.

\begin{figure}[ht]
\centering
\includegraphics[width=0.45\textwidth]{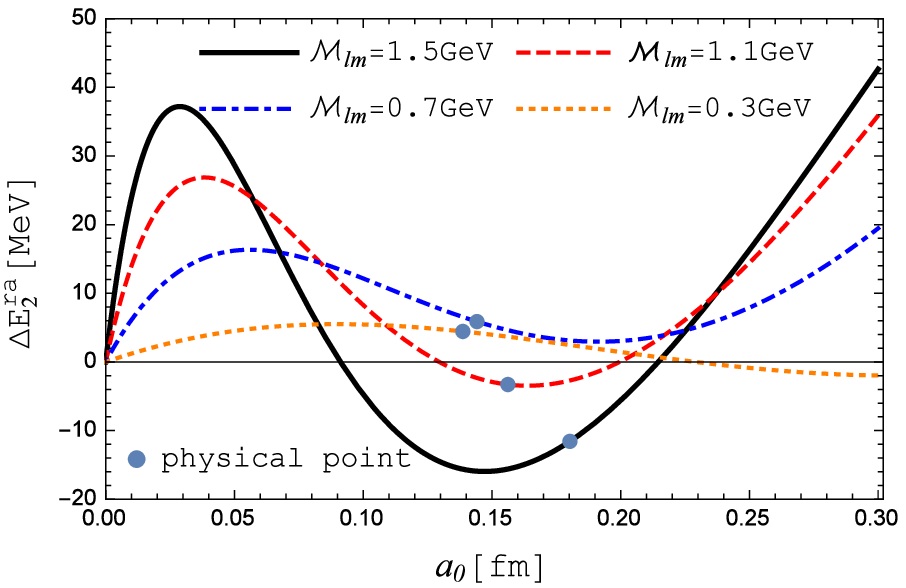}
\includegraphics[width=0.45\textwidth]{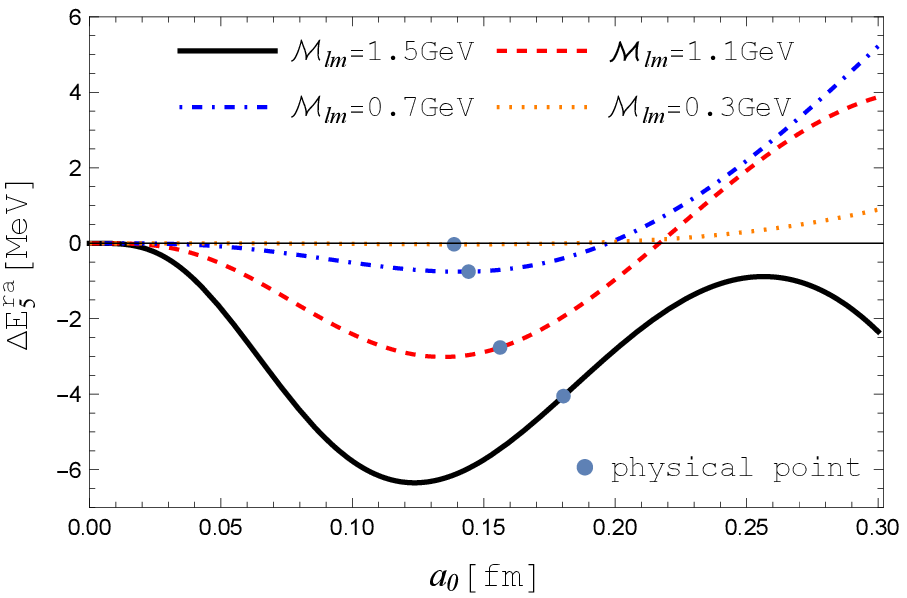}
\caption{The radial part of the perturbative corrections $\Delta E^{{\rm ra}}_2$ (the left plot) and $\Delta E^{{\rm ra}}_5$ (the right plot) as a function of the most probable radius $a_0$ at fixed ${\cal{M}}_{lm}$. The ``physical point" of $\Upsilon$ is denoted by a filled circle. }
\label{fig4}
\end{figure}

As demonstrated in the left plot of Fig.~\ref{fig4}, the magnitude of $\Delta E^{{\rm ra}}_2$ is always smaller than $\sim 40\ {\rm MeV}$ when we vary the radius at a given ${\cal{M}}_{lm}$. Numerically, we find that the maximum of $|\Delta E^{{\rm ra}}_2|$ appears as both variables equal the largest values that they may take, namely, $a_0=0.3\ {\rm fm}$ and ${\cal{M}}_{lm} = 1500\ {\rm {MeV}}$. 
As compared with our previous analysis, the upper bound of $\Delta E_2$ now is reduced to $\sim 0.3\ {\rm MeV}$. Although small enough in magnitude, we would like to take it as an overestimation. In fact, for any specific quarkonium state, in order to get $\Delta E^{{\rm ra}}_2$ at any fixed ${\cal{M}}_{lm}$, the corresponding root-mean-square radius or $a_0$ has to be determined by (numerically) solving the Schr\"odinger equation with an isotropic HQ potential $V(r, {\cal{M}}_{lm})$. According to the obtained values of the most probable radius $a_0$, one can locate a point on each curve in Fig.~\ref{fig4} which we refer to as the ``physical point". Taking the $\Upsilon$ state as an example, the ``physical point" which corresponds to the physical $a_0$ of the quarkonium state in consideration, is denoted by a filled circle in Fig.~\ref{fig4}.  

\begin{figure}[htbp]
\centering
\includegraphics[width=0.5\textwidth]{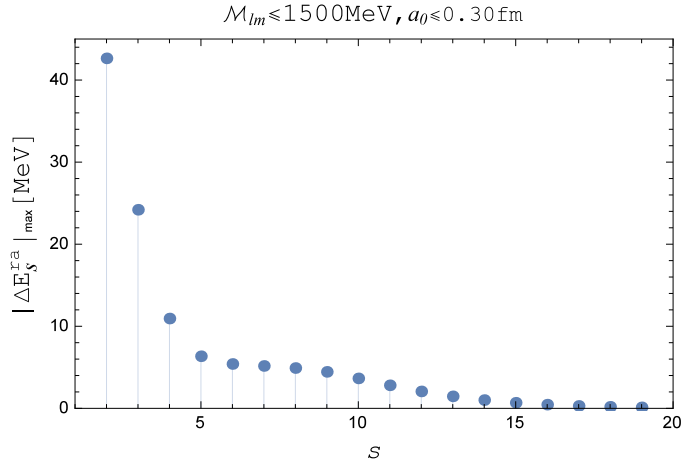}
\caption{
The maximum of $|\Delta E^{ra}_s|$ at different values of $s$ obtained under the constraint conditions ${\cal{M}}_{lm}\le 1500\,{\rm MeV}$ and $a_0 \le 0.3\,{\rm fm}$. }
\label{sdepen}
\end{figure}

In addition, under the constraint conditions that $a_0\le 0.3 \ {\rm fm}$ and ${\cal{M}}_{lm}\le 1500\ {\rm MeV}$, the maximum of $|\Delta E^{{\rm ra}}_s|$ can be determined for any given $s$. In general, the maximum appears when both $a_0$ and ${\cal{M}}_{lm}$ take their largest possible values. However, there are also exceptions such as $|\Delta E^{{\rm ra}}_5|_{\rm max}=|\Delta E^{{\rm ra}}_5 (a_0\approx 0.124\ {\rm fm},{\cal{M}}_{lm} = 1500\ {\rm MeV})|$, see the right plot of Fig.~\ref{fig4}. The corresponding results as presented in Fig.~\ref{sdepen} suggest that as compared to $|\Delta E^{{\rm ra}}_2|_{\rm max}$, higher order terms $\Delta E^{{\rm ra}}_{s>2}$ rapidly decrease in magnitude and become negligibly small even at moderate $s$. In fact, both Eq.~(\ref{dec}) and Eq.~(\ref{des}) vanish when $s$ is infinitely large. Notice that the sequential suppression of $|\Delta E^{{\rm ra}}_s|$ can not be guaranteed because the contribution from $\Delta E^{{\rm ra}}_2$ can be zero as long as certain values are assigned to $a_0$ and $ {\cal{M}}_{lm}$, see Fig.~\ref{fig4}. On the other hand, due to the strong suppression on $|\Delta E^{{\rm ra}}_s|_{\rm max}$ as well as the angular part $|\Delta E^{{\rm ag}}_s|$, the perturbative corrections $\Delta E$ as defined in Eq.~(\ref{ec}) can be neglected within an error about $\sim 0.3\  {\rm MeV}$.

According to Table~\ref{corr00}, although the differences among the eigen-energies of bottomonium obtained with various {\em effective screening mass} are not significant, the necessity of using the ``most similar state"  could be understood more clearly when we look at the energy splitting of quarkonium states with non-zero angular momentum which is a unique feature arising in an anisotropic medium. Without choosing the ``most similar state", the contribution from perturbative correction could be on the order of $\sim 1\ {\rm MeV}$ which is larger than the upper bound $\sim 0.3\  {\rm MeV}$ estimated for the ``most similar state" and actually comparable to the splitting between the states with different polarizations. For example, solving the exact three-dimensional SE, we find that the eigen-energy of $\chi_{b0}$ differs from that of $\chi_{b\pm1}$ by an amount $\sim 1 \ {\rm MeV}$. Therefore, the ``most similar state", which is expected to provide the most accurate results, is essential to reducing the three-dimensional problem effectively to an isotropic one-dimensional problem. Excellent agreement can be seen in Tables~\ref{xi1}, \ref{xi10}, and \ref{xi100} when comparing the corresponding eigen-energies with the exact solutions.

\subsection{The heavy-quark potential at large distances}

For quarkonium studies, the binding energies of the bound states are of particular interest. Therefore, the behavior of the heavy-quark potential at large distances is essentially important since the binding energy, by definition, equals the eigen-energy minus $V_{\infty}(\lambda,\xi)$. In an anisotropic plasma, $V_{\infty}(\lambda,\xi)$ is given by the following expectation values
\be\la{vinf}
V_{\infty}(\lambda,\xi)= \langle \psi_{\rm aniso}^{[k]}({\bf r}) | V(r\rightarrow \infty, \tilde{m}_D(\lambda,\xi,\theta)) | \psi_{\rm aniso}^{[k]}({\bf r})\rangle\, ,
\ee
where in principle the wave function $\psi_{\rm aniso}^{[k]}({\bf r})$ has to be determined by solving the three-dimensional SE. On the other hand, expanding $V(r\rightarrow \infty, \tilde{m}_D(\lambda,\xi,\theta))$ around the {\em effective screening mass}, we find that the leading term $V^{(0)}_{\infty}(\lambda,\xi)$ which is independent on the azimuthal angle $\theta$ turns to be an ideal approximation of $V_{\infty}(\lambda,\xi)$ when the ``most similar state" is considered.

Using Eq.~(\ref{kms}) as the heavy-quark potential model, $V_{\infty}(\lambda,\xi)$ can be expressed as the following Taylor series
\ba\la{vinfexpand}
V_{\infty}(\lambda,\xi)&\approx& \langle \psi_{nlm}({\bf r}) | V(r\rightarrow \infty, \tilde{m}_D(\lambda,\xi,\theta)) | \psi_{nlm}({\bf r})\rangle\, \nonumber \\
&=&\frac{2\sigma }{{\cal{M}}_{l m}(\lambda,\xi)}\Big[1+\sum_{s=2}^{\infty} (-1)^s\langle {\rm{Y}}_{lm}(\theta,\phi) |{\cal{F}}_{l m}^s(\xi,\theta) | {\rm{Y}}_{lm}(\theta,\phi) \rangle\Big]\, .
\ea
In the first line of the above equation, we have used the ``most similar state" denoted as $\psi_{nlm}({\bf r})$ to approximate the corresponding wave-function $\psi_{\rm aniso}^{[k]}({\bf r})$ in an anisotropic plasma. As a result, the $s=1$ term in the above sum is absent. The higher order corrections with $s\ge 2$ are suppressed as compared to the leading contribution $V^{(0)}_{\infty}(\lambda,\xi)=2\sigma /{\cal{M}}_{l m}(\lambda,\xi)$ by a small factor of $\Delta E^{{\rm ag}}_s$. In fact, we can further show the following result according to Eq.~(\ref{de2})
\be\la{de}
\frac{\Delta V_{\infty}(\lambda,\xi)}{V^{(0)}_{\infty}(\lambda,\xi)}<\sum_{s=2}^{\infty}|\Delta E^{{\rm ag}}_{s}(\xi)|<\begin{cases}
0.0055 & {\rm for} \quad l=0,m=0;\\
0.0037 &  {\rm for} \quad  l=1,m=0;\\
0.0033 &  {\rm for} \quad l=1,m=\pm1;\\
0.0094 &  {\rm for} \quad l=2,m=0,
\end{cases}
\ee
where $\Delta V_{\infty}(\lambda,\xi)\equiv V_{\infty}(\lambda,\xi)-V^{(0)}_{\infty}(\lambda,\xi)$.

Because the heavy-quark potential at finite temperature is deeper than the vacuum potential, one can expect that $V_{\infty}(\lambda,\xi)$ and $V^{(0)}_{\infty}(\lambda,\xi)$ are smaller than $V_{\infty}(\lambda=0)$. Roughly speaking, $V_{\infty}(\lambda=0)$ corresponds to the value of the vacuum potential at the string breaking distance which, in general, is approximately $1\ {\rm GeV}$.  Therefore, one can estimate that the higher order correction $\Delta V_{\infty}(\lambda,\xi) < 10 \ {\rm MeV}$.

\begin{center}
\begin{table}[htpb]
\setlength{\tabcolsep}{7mm}{
\begin{tabular}{  c  c  c  c   c}
\toprule[1pt]
$                        $ & $V^{(0)}_{\infty}$ & $V^{(1)}_{\infty}$ & $V^{(2)}_{\infty}$ & $    V^{(3)}_{\infty}      $      \\ \hline
$ {\cal {M}}_{00}$ & $894.24$  & $ 0             $ & $ 1.86       $ & $0.0338 $      \\
${\cal {M}}_{10}$  & $932.06$  & $-39.41         $ & $ 3.77       $ & $-0.297$      \\
${\cal {M}}_{11}$  & $876.46$  & $17.43        $ & $ 2.09       $ & $0.142  $      \\ 	
${\cal {M}}_{20}$  & $917.43$  & $-23.79         $ & $ 2.62       $ & $-0.135 $      \\
${\cal {M}}_{21}$  & $909.67$  & $-15.69        $ & $2.23        $ & $-0.070  $      \\
\bottomrule[1pt]	
\end{tabular}
}
\caption{The values of $V^{(0)}_{\infty}(\lambda,\xi)$ and their perturbative corrections for the ground state bottomonium evaluated with different  {\em effective screening masses}.  The results are obtained at $\xi=1$, $\sigma=0.223\  {\rm GeV}^2$ and $\lambda=1.1 \, T_c$ with $T_c=192 \ {\rm MeV}$.  All results are given in the units of {\rm {MeV}}.}
\label{vinf00}
\end{table}
\end{center}

In addition, the necessity of using the ``most similar state"  could be further justified when we expand $V(r\rightarrow \infty, \tilde{m}_D(\lambda,\theta,\xi))$ around the {\em effective screening mass} ${\cal{M}}_{l^\prime m^\prime}(\lambda,\xi)$ with different quantum numbers $l^\prime$ and $m^\prime$. The leading term $V^{(0)}_{\infty}(\lambda,\xi)=2\sigma /{\cal{M}}_{l^\prime m^\prime}(\lambda,\xi)$ while the corrections can be denoted as
\be
V^{(s)}_{\infty}(\lambda,\xi)=\frac{2\sigma }{{\cal{M}}_{l^\prime m^\prime}(\lambda,\xi)}(-1)^s\langle {\rm{Y}}_{l m}(\theta,\phi) |{\cal{F}}_{l^\prime m^\prime}^s(\xi,\theta) | {\rm{Y}}_{lm}(\theta,\phi) \rangle\, .
\ee
Taking $(l,m)=(0,0)$ as an example, the Taylor series of $V_{\infty}(\lambda,\xi)$ has been computed and the results are shown in Table~\ref{vinf00}. The ``most similar state" corresponds to using the {\em effective screening mass} ${\cal{M}}_{00}(\lambda,\xi)$.  In this case, $V^{(0)}_{\infty}(\lambda,\xi)$ turns out to be a good approximation to $V_{\infty}(\lambda,\xi)$ which is $896.20\,{\rm MeV}$ from the exact 3D evaluation. On the other hand, with other {\em effective screening mass} determined with $(l^\prime m^\prime)\neq (0,0)$, the non-vanishing $V^{(1)}_{\infty}(\lambda,\xi)$ may give rise to a considerable correction on the binding energy, especially for excited states. More importantly, based on the estimation in Ref.~\cite{Dumitru:2009ni}, the splitting of the binding energy of $\chi_b$ with different angular momentum is on the order of $50\,{\rm MeV}$ which is obviously comparable with the contribution from $V^{(1)}_{\infty}(\lambda,\xi)$. As a result, when keeping only the leading term in Eq.~(\ref{vinfexpand}), one has to make use of the ``most similar state" in order to get a quantitatively reliable result.

\section{Some applications}\la{apps}

\begin{figure}[!htbp]
\centering
\includegraphics[width=0.5\textwidth]{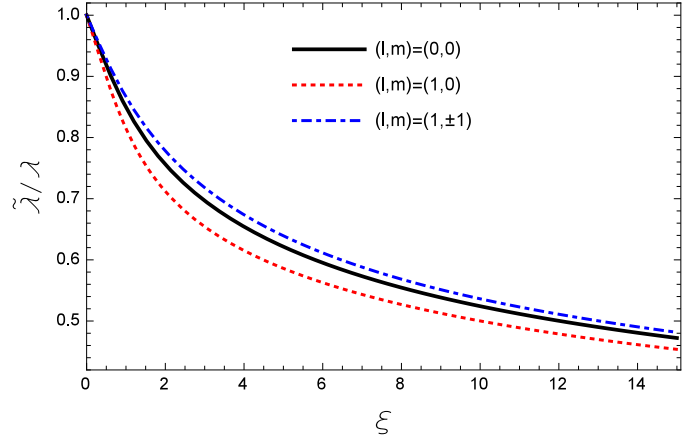}
\vspace*{-0.4cm}
\caption{
The ratio ${\tilde \lambda}/\lambda$ as a function of $\xi$ at some fixed $(l,m)$.}
\label{etem}
\end{figure}

Perturbatively, the Debye mass $m_D$ at leading order is proportional to the temperature $\lambda$ or $T$ in an isotropic (equilibrium) medium. On the other hand, studies on quarkonium states become most relevant in a temperature region not from far above $T_c$ where non-perturbative physics plays an important role. As argued in Ref.~\cite{Kaczmarek:2005ui}, lattice simulations suggest that the non-perturbative Debye mass can be approximated as a constant factor $A$ times the perturbative $m_D$, therefore, as used in our numerical evaluations, $m_D$ is also proportional to $\lambda$ at relatively lower temperatures, roughly speaking, from $T_c$ to $\sim 3 T_c$, which is known as the semi-QGP \cite{Dumitru:2010mj,Dumitru:2012fw,Guo:2014zra,Pisarski:2016ixt}. With the {\em effective screening mass} ${\cal{M}}_{l m}(\lambda,\xi)$ given in Eq.~(\ref{effm0}), it is possible to define an effective temperature ${\tilde \lambda}$ in an anisotropic medium by which the {\em effective screening mass} can be formally expressed as ${\cal{M}}_{l m}(\lambda,\xi)=A g {\tilde \lambda} \sqrt{1+N_f/6}$. The effective temperature ${\tilde \lambda}$ is determined by requiring that the screening mass ${\cal{M}}_{l m}(\lambda,\xi)$ or the binding energy of a bound state in an anisotropic medium is equal to that in an isotropic medium determined at temperature ${\tilde \lambda}$. Apparently, ${\tilde \lambda}$ depends not only on the anisotropy $\xi$ but also on the quantum numbers $l$ and $m$. According to Eqs.~(\ref{eff11}) and (\ref{effm0}), we can show that
\be
\frac{{\tilde \lambda}}{\lambda}=\int_{-1}^{1} d \cos \theta \int_{0}^{2\pi} d \phi {\rm{Y}}_{l m}(\theta,\phi)  \big[f_1(\xi) \cos^2 \theta+f_2(\xi)\big]^{-\frac{1}{4}} {\rm{Y}}^*_{l m}(\theta,\phi)\, .
\ee
The ratio ${\tilde \lambda}/\lambda$ as a function of $\xi$ at some fixed $(l,m)$ is shown in Fig.~\ref{etem}. As we can see, a linear decrease with the increasing anisotropies appears in the small $\xi$ region. This is actually in accordance with Eq.~(\ref{effm0small}) by which the above ratio reduces to $1- k(l,m) \xi/8$ for $\xi \ll 1$.

The perturbative Debye mass at non-zero quark chemical potential $\mu$ is given by
\be\la{mdu}
m_D= g \lambda \sqrt{1+N_f\Big(\frac{1}{6}+2 {\tilde{\mu}}^2\Big)}\,
\ee
with ${\tilde{\mu}}\equiv \mu/(2\pi \lambda)$, which suggests an enhanced screening in the high temperature limit where the potential takes a Debye-screened form. Therefore, the chemical potential affects the binding of the bound states in an opposite way as compared to the momentum space anisotropies. In the semi-QGP region, assuming the potential model is formally unchanged when introducing a chemical potential and the corresponding $\mu$-modifications can be entirely encoded into the Debye mass in exactly the same way as the perturbative case, Eq.~(\ref{mdu}), then the  {\em effective screening mass} at non-zero chemical potential reads
\be
 {\cal{M}}_{l m}(\lambda,\xi, \mu)={\cal{M}}_{l m}(\lambda,\xi, \mu=0)  \sqrt{1+2 N_f  {\tilde{\mu}}^2/(1+N_f/6)}\, .
 \ee
As a result, one can consider the competition between the two different effects on the binding of a bound state, namely, the anisotropies which lead to a tightly bound quarkonium state and the chemical potential which decreases the binding. In particular, a complete cancellation between the two effects happens when ${\cal{M}}_{l m}(\lambda,\xi,\mu)={\cal{M}}_{l m}(\lambda,0,0)=A g \lambda \sqrt{1+N_f/6}$. In Fig.~\ref{app2}, the curve on the ${\tilde{\mu}} -\xi$ plane indicates such a complete cancellation which, in the small $\xi$ limit, corresponds to ${\tilde{\mu}}/\sqrt{\xi}=\sqrt{k(l,m) (N_f+6)/(48N_f)} $.

\begin{figure}
\centering
\includegraphics[width=0.5\textwidth]{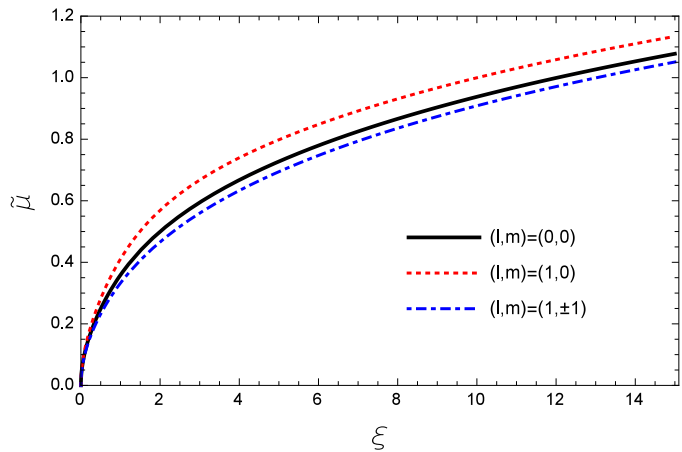}
\vspace*{-0.3cm}
\caption{The relation between ${\tilde{\mu}}$ and $\xi$. At a given temperature $\lambda$, the screening in an anisotropic medium with $\xi$ and $\mu$ satisfying this relation is identical to that in an isotropic medium with $\mu=0$. In this plot, we choose $N_f=2$.}
\la{app2}
\end{figure}

Finally, as mentioned in the introduction, one of the primary motivations for this study was to assess whether it is possible to have an effectively isotropic model potential that reproduces the binding energies of different quarkonium states in a momentum-anisotropic QGP. Our findings can be used to include the effect of momentum anisotropies on in-medium bound state evolution using real-time solution of the Schr\"odinger equation in a complex screened potential, see e.g. \cite{Islam:2020gdv,Islam:2020bnp}.  However, to do this properly one must prove that the same logic used herein for the real part of the potential can be applied to the imaginary part of the heavy-quark potential.  Our preliminary results \cite{aniso2}, indicate that the same method can be applied to achieve a numerically reliable isotropic model of the imaginary part of the potential as well.  Once this step is complete it will be possible to assess momentum-space anisotropy effects on heavy-quarkonium using isotropic (effectively 1D) simulations.

\section{Conclusions and Outlook}\la{conclusions}

In this paper, we introduced a prescription for an isotropic effective Deybe mass that depends on the quantum numbers $l$ and $m$ of heavy-quarkonium state.  This mass is obtained by integrating the angle-dependent Debye mass, which emerges when $\xi \neq 0$, with spherical harmonic basis functions \eqref{effm0}.  Tables \ref{xi1}-\ref{xi100} in the appendix demonstrate that when using an isotropic potential with \eqref{effm0} as the effective Debye mass, we can reproduce the energy and binding energies obtained by direct numerical solution of the 3D Schr\"odinger equation for the same underlying anisotropic potential.  Our results demonstrate that, for both small and large anisotropy parameters, one can reproduce the energy to within fractions of an MeV and the binding energy to within a few MeV.  As these tables also demonstrate, with this method we are even able to resolve the splitting of the different $p$-wave states in an anisotropic potential model.

After introducing this method, we discussed why one expects this to be a good approximation even when there is a high degree of momentum anisotropy.  We demonstrated that higher-order corrections are under control and that the resulting series converge very quickly, which explains why this prescription does so well in numerically reproducing the full 3D results.  Following this, we mentioned some applications of the method introduced herein which include using in real-time solution of the Sch\"odinger equation with a complex in-medium potential.  To complete this task, we are now investigating if similar techniques can be applied to the imaginary part of the heavy-quark potential.  Preliminary results in this direction are quite promising \cite{aniso2}.

\section*{Acknowledgements}
The work of Y.G. is supported by the NSFC of China under Project Nos. 11665008 and 12065004. M.S. and A.I.  were supported by the U.S. Department of Energy, Office of Science, Office of Nuclear Physics Award No.~DE-SC0013470.

\appendix

\section{Tables}\label{appa}
In this appendix, by numerically solving the 3D Schr\"odinger equation with a HQ potential given in Eq.~(\ref{kms}), we list the exact results of the eigen-energies ($E$) and binding-energies ($E_B$) for several low-lying heavy-quarkonium bound states, including $\Upsilon$, $\chi_{b0}$, $\chi_{b \pm1}$ as well as $J/\Psi$. We consider various temperatures relevant for quarkonium studies with small ($\xi=1$), moderate ($\xi=10$) and large ($\xi=100$) anisotropies. Comparing with the results obtained based on the one-dimensional potential model with {\em effective screening masses}, the corresponding discrepancies as denoted by $\delta E$ and $\delta E_B$ are also listed for directly testing our method.  For the 3D code, we used a previously developed code called quantumFDTD \cite{Dumitru:2009ni,Strickland:2009ft,Margotta:2011ta,QFDTD2,Strickland:2011aa,Delgado:2020ozh}.

\begin{center}
\begin{table}[htpb]
\setlength{\tabcolsep}{7mm}{
\begin{tabular}{  c  c  c  c c }
\toprule[1pt]
$    \Upsilon       $  & $    E      $   & $  \delta E  $ & $     E_B      $ & $ \delta E_B $      \\ \hline
$       T_c           $  & $112.568$   & $0.038        $ & $-873.247   $ & $2.112          $     \\
$    1.1T_c         $  & $115.358$   & $0.044        $ & $-780.843   $ & $1.917          $      \\
$    1.2T_c         $  & $118.237$   & $0.039        $ & $-703.285   $ & $1.763          $      \\
$    1.4T_c         $  & $124.182$   & $0.040        $ & $-579.985   $ & $1.509          $      \\
$    1.6T_c         $  & $130.257$   & $0.035        $ & $-485.890   $ & $1.322          $      \\
$    1.8T_c         $  & $136.341$   & $0.038        $ & $-411.345   $ & $1.167          $      \\
$     2.0T_c        $  & $142.328$   & $0.036        $ & $-350.585   $ & $1.044          $      \\
\bottomrule[1pt]	
\end{tabular}
\vspace{0.2cm}
\vspace{0.2cm}
\begin{tabular}{  c  c  c  c c }
\toprule[1pt]
$    \chi_{b0}      $  & $    E      $   & $  \delta E  $ & $     E_B      $ & $ \delta E_B $      \\ \hline
$       T_c           $  & $509.811$   & $0.088        $ & $-516.631   $ & $1.270          $     \\
$    1.1T_c         $  & $507.878$   & $0.091        $ & $-425.186   $ & $1.100          $      \\
$    1.2T_c         $  & $505.564$   & $0.078        $ & $-349.671   $ & $0.930          $      \\
$    1.4T_c         $  & $499.672$   & $0.042        $ & $-233.215   $ & $0.600          $      \\
$    1.6T_c         $  & $491.906$   & $0.066        $ & $-149.142   $ & $0.195          $      \\
$    1.8T_c         $  & $481.706$   & $0.020        $ & $-87.804     $ & $0.099          $      \\
$     2.0T_c        $  & $468.857$   & $0.042        $ & $-43.204     $ & $0.611          $      \\
\bottomrule[1pt]	
\end{tabular}
\vspace{0.2cm}
\vspace{0.2cm}
\begin{tabular}{  c  c  c  c c }
\toprule[1pt]
$ \chi_{b\pm1}   $  & $    E      $   & $  \delta E  $ & $     E_B      $ & $ \delta E_B $      \\ \hline
$       T_c           $  & $508.635$   & $0.082        $ & $-456.430   $ & $1.039          $     \\
$    1.1T_c         $  & $506.318$   & $0.072        $ & $-370.959   $ & $0.887          $      \\
$    1.2T_c         $  & $503.547$   & $0.050        $ & $-300.562   $ & $0.736          $      \\
$    1.4T_c         $  & $496.510$   & $0.031        $ & $-192.593   $ & $0.423          $      \\
$    1.6T_c         $  & $487.003$   & $0.007        $ & $-115.759   $ & $0.188          $      \\
$    1.8T_c         $  & $474.756$   & $0.044        $ & $-60.749     $ & $0.154          $      \\
$     2.0T_c        $  & $458.935$   & $0.023        $ & $-22.513     $ & $0.629          $      \\
\bottomrule[1pt]	
\end{tabular}
\vspace{0.2cm}
\vspace{0.2cm}
\begin{tabular}{  c  c  c  c c }
\toprule[1pt]
$      J/\Psi         $  & $    E      $   & $  \delta E  $ & $     E_B      $ & $ \delta E_B $      \\ \hline
$       T_c           $  & $464.659$   & $0.013        $ & $-520.724   $ & $1.732          $     \\
$    1.1T_c         $  & $460.629$   & $0.015        $ & $-435.095   $ & $1.498          $      \\
$    1.2T_c         $  & $456.190$   & $0.021        $ & $-364.807   $ & $1.298          $      \\
$    1.4T_c         $  & $446.067$   & $0.035        $ & $-257.468   $ & $0.953          $      \\
$    1.6T_c         $  & $434.287$   & $0.028        $ & $-181.103   $ & $0.627          $      \\
$    1.8T_c         $  & $420.917$   & $0.038        $ & $-125.865   $ & $0.264          $      \\
$     2.0T_c        $  & $405.852$   & $0.005        $ & $-85.939     $ & $0.037          $      \\

\bottomrule[1pt]	
\end{tabular}
}
\caption{The exact results of the eigen-energies ($E$) and binding-energies ($E_B$) for different quarkonium states at various temperatures with $\xi=1$. Comparing with the results obtained based on the one-dimensional potential model with  {\em effective screening masses}, the corresponding discrepancies as denoted by $\delta E$ and $\delta E_B$ are also listed. All the results are given in the unit of ${\rm {MeV}}$.}
\label{xi1}
\end{table}
\end{center}

\begin{center}
\begin{table}[htpb]
\setlength{\tabcolsep}{7mm}{
\begin{tabular}{  c  c  c  c c }
\toprule[1pt]
$    \Upsilon       $  & $    E      $   & $  \delta E  $ & $     E_B      $ & $ \delta E_B $      \\ \hline
$       T_c           $  & $103.145$   & $0.038        $ & $-1494.920   $ & $4.404          $     \\
$    1.1T_c         $  & $104.494$   & $0.044        $ & $-1348.302   $ & $4.004          $      \\
$    1.2T_c         $  & $105.917$   & $0.040        $ & $-1225.821   $ & $3.678          $      \\
$    1.4T_c         $  & $108.964$   & $0.043        $ & $-1032.540   $ & $3.159          $      \\
$    1.6T_c         $  & $112.232$   & $0.046        $ & $-886.595   $ & $2.767          $      \\
$    1.8T_c         $  & $115.675$   & $0.040        $ & $-772.181   $ & $2.470          $      \\
$     2.0T_c        $  & $119.250$   & $0.045        $ & $-679.827   $ & $2.221          $      \\
\bottomrule[1pt]	
\end{tabular}
\vspace{0.2cm}
\vspace{0.2cm}
\begin{tabular}{  c  c  c  c c }
\toprule[1pt]
$    \chi_{b0}      $  & $    E      $   & $  \delta E  $ & $     E_B        $ & $ \delta E_B $      \\ \hline
$       T_c           $  & $514.280$   & $0.098        $ & $-1159.014   $ & $2.886          $     \\
$    1.1T_c         $  & $513.840$   & $0.095        $ & $-1007.302   $ & $2.595          $      \\
$    1.2T_c         $  & $513.309$   & $0.100        $ & $-881.035     $ & $2.355          $      \\
$    1.4T_c         $  & $511.942$   & $0.101        $ & $-683.133     $ & $1.957          $      \\
$    1.6T_c         $  & $510.115$   & $0.097        $ & $-535.489     $ & $1.635          $      \\
$    1.8T_c         $  & $507.770$   & $0.092        $ & $-421.557     $ & $1.360          $      \\
$     2.0T_c        $  & $504.847$   & $0.085        $ & $-331.432     $ & $1.111          $      \\
\bottomrule[1pt]	
\end{tabular}
\vspace{0.2cm}
\vspace{0.2cm}
\begin{tabular}{  c  c  c  c c }
\toprule[1pt]
$ \chi_{b\pm1}   $  & $    E      $   & $  \delta E  $ & $     E_B        $ & $ \delta E_B $      \\ \hline
$       T_c           $  & $513.972$   & $0.100        $ & $-1046.152   $ & $2.448          $     \\
$    1.1T_c         $  & $513.427$   & $0.096        $ & $-904.835     $ & $2.198          $      \\
$    1.2T_c         $  & $512.771$   & $0.096        $ & $-787.268     $ & $1.988          $      \\
$    1.4T_c         $  & $511.083$   & $0.090        $ & $-603.165     $ & $1.640          $      \\
$    1.6T_c         $  & $508.833$   & $0.090        $ & $-466.053     $ & $1.365          $      \\
$    1.8T_c         $  & $505.949$   & $0.074        $ & $-360.526     $ & $1.118          $      \\
$     2.0T_c        $  & $502.364$   & $0.051        $ & $-277.363     $ & $0.890          $      \\
\bottomrule[1pt]	
\end{tabular}
\vspace{0.2cm}
\vspace{0.2cm}
\begin{tabular}{  c  c  c  c c }
\toprule[1pt]
$      J/\Psi         $  & $    E      $   & $  \delta E  $ & $     E_B        $ & $ \delta E_B $      \\ \hline
$       T_c           $  & $476.428$   & $0.007        $ & $-1121.278   $ & $4.076          $     \\
$    1.1T_c         $  & $474.908$   & $0.005        $ & $-977.498     $ & $3.652          $      \\
$    1.2T_c         $  & $473.245$   & $0.003        $ & $-858.073     $ & $3.301          $      \\
$    1.4T_c         $  & $469.481$   & $0.013        $ & $-671.539     $ & $2.730          $      \\
$    1.6T_c         $  & $465.120$   & $0.017        $ & $-533.157     $ & $2.280          $      \\
$    1.8T_c         $  & $460.146$   & $0.032        $ & $-427.088     $ & $1.920          $      \\
$     2.0T_c        $  & $454.546$   & $0.040        $ & $-343.832     $ & $1.606          $      \\

\bottomrule[1pt]	
\end{tabular}
}
\caption{The exact results of the eigen-energies ($E$) and binding-energies ($E_B$) for different quarkonium states at various temperatures with $\xi=10$. Comparing with the results obtained based on the one-dimensional potential model with  {\em effective screening masses}, the corresponding discrepancies as denoted by $\delta E$ and $\delta E_B$ are also listed. All the results are given in the unit of ${\rm {MeV}}$.}
\label{xi10}
\end{table}
\end{center}

\begin{center}
\begin{table}[htpb]
\setlength{\tabcolsep}{7mm}{
\begin{tabular}{  c  c  c  c c }
\toprule[1pt]
$    \Upsilon       $  & $    E      $   & $  \delta E  $ & $     E_B      $ & $ \delta E_B $      \\ \hline
$       T_c           $  & $98.209$   & $0.030        $ & $-2808.832   $ & $1.042          $     \\
$    1.1T_c         $  & $98.691$   & $0.031        $ & $-2544.075   $ & $0.945          $      \\
$    1.2T_c         $  & $99.208$   & $0.030        $ & $-2323.329   $ & $0.867          $      \\
$    1.4T_c         $  & $100.341$   & $0.026        $ & $-1976.120   $ & $0.743          $      \\
$    1.6T_c         $  & $101.599$   & $0.033        $ & $-1715.307   $ & $0.643          $      \\
$    1.8T_c         $  & $102.968$   & $0.028        $ & $-1512.061   $ & $0.574          $      \\
$     2.0T_c        $  & $104.439$   & $0.027        $ & $-1349.089   $ & $0.517          $      \\
\bottomrule[1pt]	
\end{tabular}
\vspace{0.2cm}
\vspace{0.2cm}
\begin{tabular}{  c  c  c  c c }
\toprule[1pt]
$    \chi_{b0}      $  & $    E      $   & $  \delta E  $ & $     E_B          $ & $ \delta E_B $      \\ \hline
$       T_c           $  & $515.341$   & $0.097        $ & $-2441.733     $ & $0.863          $     \\
$    1.1T_c         $  & $515.267$   & $0.102        $ & $-2172.979     $ & $0.796          $      \\
$    1.2T_c         $  & $515.178$   & $0.100        $ & $-1949.046     $ & $0.734          $      \\
$    1.4T_c         $  & $514.944$   & $0.097        $ & $-1597.243     $ & $0.636          $      \\
$    1.6T_c         $  & $514.627$   & $0.094        $ & $-1333.531     $ & $0.559          $      \\
$    1.8T_c         $  & $514.216$   & $0.093        $ & $-1128.586     $ & $0.502          $      \\
$     2.0T_c        $  & $513.697$   & $0.093        $ & $-964.818       $ & $0.454          $      \\
\bottomrule[1pt]	
\end{tabular}
\vspace{0.2cm}
\vspace{0.2cm}
\begin{tabular}{  c  c  c  c c }
\toprule[1pt]
$ \chi_{b\pm1}   $  & $    E      $   & $  \delta E  $ & $     E_B        $ & $ \delta E_B $      \\ \hline
$       T_c           $  & $515.324$   & $0.099        $ & $-2366.677   $ & $0.666          $     \\
$    1.1T_c         $  & $515.244$   & $0.095        $ & $-2104.755   $ & $0.609          $      \\
$    1.2T_c         $  & $515.146$   & $0.101        $ & $-1886.518   $ & $0.571          $      \\
$    1.4T_c         $  & $514.893$   & $0.095        $ & $-1543.672   $ & $0.493          $      \\
$    1.6T_c         $  & $514.550$   & $0.095        $ & $-1286.690   $ & $0.439          $      \\
$    1.8T_c         $  & $514.105$   & $0.089        $ & $-1086.994   $ & $0.392          $      \\
$     2.0T_c        $  & $513.544$   & $0.094        $ & $-927.441     $ & $0.362          $      \\
\bottomrule[1pt]	
\end{tabular}
\vspace{0.2cm}
\vspace{0.2cm}
\begin{tabular}{  c  c  c  c c }
\toprule[1pt]
$      J/\Psi         $  & $    E       $   & $  \delta E  $ & $     E_B         $ & $ \delta E_B $      \\ \hline
$       T_c           $  & $481.569$   & $0.021        $ & $-2425.444     $ & $1.024          $     \\
$    1.1T_c         $  & $481.095$   & $0.020        $ & $-2161.640     $ & $0.926          $      \\
$    1.2T_c         $  & $480.580$   & $0.016        $ & $-1941.923     $ & $0.847          $      \\
$    1.4T_c         $  & $479.427$   & $0.015        $ & $-1596.996     $ & $0.716          $      \\
$    1.6T_c         $  & $478.111$   & $0.019         $ & $-1338.752     $ & $0.614          $      \\
$    1.8T_c         $  & $476.628$   & $0.016         $ & $-1138.353     $ & $0.538          $      \\
$     2.0T_c        $  & $474.976$   & $0.019         $ & $-978.499       $ & $0.471          $      \\

\bottomrule[1pt]	
\end{tabular}
}
\caption{The exact results of the eigen-energies ($E$) and binding-energies ($E_B$) for different quarkonium states at various temperatures with $\xi=100$. Comparing with the results obtained based on the one-dimensional potential model with  {\em effective screening masses}, the corresponding discrepancies as denoted by $\delta E$ and $\delta E_B$ are also listed. All the results are given in the unit of ${\rm {MeV}}$.}
\label{xi100}
\end{table}
\end{center}

\end{document}